\documentclass[a4paper,dvips,12pt]{article}
\usepackage{cite}
\usepackage{epsf}
\usepackage{multirow}
\usepackage{bm,bbm} 
\usepackage[pdftex]{graphicx}
\usepackage{tabularx}
\usepackage{latexsym}
\usepackage{amsmath,amssymb,exscale}
\usepackage{array,multicol}
\numberwithin{equation}{section}

\def\id{\ 1 \! \! \! \! 1}
\def\draftlabel#1{{\@bsphack\if@filesw {\let\thepage\relax
   \xdef\@gtempa{\write\@auxout{\string
      \newlabel{#1}{{\@currentlabel}{\thepage}}}}}\@gtempa
   \if@nobreak \ifvmode\nobreak\fi\fi\fi\@esphack}
        \gdef\@eqnlabel{#1}}
\def\@eqnlabel{}
\def\@vacuum{}
\def\draftmarginnote#1{\marginpar{\raggedright\scriptsize\tt#1}}
\def\draft{\oddsidemargin -.5truein
        \def\@oddfoot{\sl preliminary draft \hfil
        \rm\thepage\hfil\sl\today\quad\militarytime}
        \let\@evenfoot\@oddfoot \overfullrule 3pt
        \let\label=\draftlabel
        \let\marginnote=\draftmarginnote
   \def\@eqnnum{(\theequation)\rlap{\kern\marginparsep\tt\@eqnlabel}%
\global\let\@eqnlabel\@vacuum}  }

\newcommand{\PRL}[3]{\emph{ Phys.~Rev.~Lett.} \textbf{#1} (#2) #3}

\newcommand{\PR}[3]{\emph{ Phys.~Rep.} \textbf{#1} (#2) #3}

\def\ov{\overline}

\def\dalemb#1#2{{\vbox{\hrule height .#2pt
         \hbox{\vrule width.#2pt height#1pt \kern#1pt
                 \vrule width.#2pt}
         \hrule height.#2pt}}}

\def\half{{\textstyle{1\over2}}}
\let\a=\alpha    
    
    \let\p=\pi 
\let\s=\sigma     

      \let\G=\Gamma  
  \let\S=\Sigma   
\let\F=\Phi

 \def\bd{\begin{document}} \def\ed{\end{document}}
\def\ds{\documentstyle} \let\fr=\frac \let\bl=\bigl \let\br=\bigr
\let\Br=\Bigr \let\Bl=\Bigl
\let\bm=\bibitem
\let\na=\nabla
\let\pa=\partial
\let\ov=\overline
\def\ie{{\it i.e.\ }}
\def\tr{{\mbox{\rm tr}}}
\newcommand{\be}{\begin{equation}}
\newcommand{\ee}{\end{equation}}
\newcommand{\beba}{\begin{equation}\begin{array}{lcl}}
\newcommand{\eaee}{\end{array}\end{equation}}
\newcommand{\bea}{\begin{eqnarray}}
\newcommand{\eea}{\end{eqnarray}}
\newcommand{\ba}{\begin{array}}
\newcommand{\ea}{\end{array}}
\newcommand{\td}{\tilde}
\newcommand{\norsl}{\normalsize\sl}
\newcommand{\ns}{\normalsize}
\newcommand{\refs}[1]{(\ref{#1})}
\def\simlt{\mathrel{\lower2.5pt\vbox{\lineskip=0pt\baselineskip=0pt
            \hbox{$<$}\hbox{$\sim$}}}}
\def\simgt{\mathrel{\lower2.5pt\vbox{\lineskip=0pt\baselineskip=0pt
            \hbox{$>$}\hbox{$\sim$}}}}
\def\A{{\cal A}}
\def\a{{\mathcal a}}
\def\V{{\cal V}}
\def\F{{\cal F}}
\def\p{{\mathcal \phi}}
\def\L{{\mathcal L}}
\def\M{{\mathcal M}}
\def\bD{{\ov {\rm D}}}
\def\bO{{\ov {\rm O}}}
\def\bOp{{\ov {\rm O'}}}
\def\O{{ {\rm O}}}

\newcommand{\nsect}{\setcounter{equation}{0}
\def\theequation{\thesection.\arabic{equation}}\section}
\newcommand{\nappend}{\setcounter{equation}{0}
\def\theequation{\rm{A}.\arabic{equation}}\section*}
\newcommand{\appendixA}{\setcounter{equation}{0}
\def\theequation{\rm{A}.\arabic{equation}}\section*}
\newcommand{\appendixB}{\setcounter{equation}{0}
\def\theequation{\rm{B}.\arabic{equation}}\section*}
\newcommand{\appendixC}{\setcounter{equation}{0}
\def\theequation{\rm{C}.\arabic{equation}}\section*}
\newcommand{\appendixD}{\setcounter{equation}{0}
\def\theequation{\rm{D}.\arabic{equation}}\section*}
\newcommand{\appendixE}{\setcounter{equation}{0}
\def\theequation{\rm{E}.\arabic{equation}}\section*}
\newcommand{\appendixF}{\setcounter{equation}{0}
\def\theequation{\rm{F}.\arabic{equation}}\section*}
\newcommand{\appendixG}{\setcounter{equation}{0}
\def\theequation{\rm{G}.\arabic{equation}}\section*}
\def\baselinestretch{1.5}
\catcode`\@=11
\def\marginnote#1{}
\newcount\hour
\newcount\minute
\newtoks\amorpm
\hour=\time\divide\hour by60
\minute=\time{\multiply\hour by60 \global\advance\minute by-\hour}
\edef\standardtime{{\ifnum\hour<12 \global\amorpm={am}%
        \else\global\amorpm={pm}\advance\hour by-12 \fi
        \ifnum\hour=0 \hour=12 \fi
        \number\hour:\ifnum\minute<10 0\fi\number\minute\the\amorpm}}
\edef\militarytime{\number\hour:\ifnum\minute<10 0\fi\number\minute}
\def\draftlabel#1{{\@bsphack\if@filesw {\let\thepage\relax
   \xdef\@gtempa{\write\@auxout{\string
      \newlabel{#1}{{\@currentlabel}{\thepage}}}}}\@gtempa
   \if@nobreak \ifvmode\nobreak\fi\fi\fi\@esphack}
        \gdef\@eqnlabel{#1}}
\def\@eqnlabel{}
\def\@vacuum{}
\def\draftmarginnote#1{\marginpar{\raggedright\scriptsize\tt#1}}
\def\draft{\oddsidemargin -.5truein
        \def\@oddfoot{\sl preliminary draft \hfil
        \rm\thepage\hfil\sl\today\quad\militarytime}
        \let\@evenfoot\@oddfoot \overfullrule 3pt
        \let\label=\draftlabel
        \let\marginnote=\draftmarginnote
   \def\@eqnnum{(\theequation)\rlap{\kern\marginparsep\tt\@eqnlabel}%
\global\let\@eqnlabel\@vacuum}  }

\def\preprint{\twocolumn\sloppy\flushbottom\parindent 1em
        \leftmargini 2em\leftmarginv .5em\leftmarginvi .5em
        \oddsidemargin -.5in    \evensidemargin -.5in
        \columnsep 15mm \footheight 0pt
        \textwidth 250mmin      \topmargin  -.4in
        \headheight 12pt \topskip .4in
        \textheight 175mm
        \footskip 0pt
\def\@oddhead{\thepage\hfil\addtocounter{page}{1}\thepage}
        \let\@evenhead\@oddhead \def\@oddfoot{} \def\@evenfoot{} }

\def\titlepage{\@restonecolfalse\if@twocolumn\@restonecoltrue\onecolumn
     \else \newpage \fi \thispagestyle{empty}\c@page\z@ 
        \def\thefootnote{\fnsymbol{footnote}} }

\def\endtitlepage{\if@restonecol\twocolumn \else  \fi
        \def\thefootnote{\arabic{footnote}}
        \setcounter{footnote}{0}}  

\catcode`@=12
\relax
\def\abs#1{\left| #1\right|}
\def\bC{\mathop{\bf C}}
\def\bea{\begin{array}}
\def\bem{\begin{displaymath}}
\def\beq{\begin{equation}}
\def\bea{\begin{eqnarray}}
\def\bR{\mathop{\bf R}}
\def\bra#1{\left\langle #1\right|}
\def\eea{\end{array}}
\def\eem{\end{displaymath}}
\def\eeq{\end{equation}}
\def\eea{\end{eqnarray}}
\def\eq{\beq\eeq}                          
\def\eqr#1{\beq\label#1\eeq}               
\def\half{\frac{1}{2}}
\def\Im{\mathop{\rm Im}}
\def\ket#1{\left| #1\right\rangle}
\def\sket#1{| #1 >}
\def\lie{\hbox{\it \$}}                          
\def\lineint{\oint \frac{d z}{2 \pi i}} 
\def\modsq#1{| #1 |^2}
\def\NP#1#2#3{Nucl. Phys. \underline{#1} (19#2) #3}
\def\ov{\overline}
\def\partder#1#2{{\partial #1\over\partial #2}}
\def\PL#1#2#3{Phys. Lett. \underline{#1} (19#2) #3}
\def\PR#1#2#3{Phys. Rev. \underline{#1} (19#2) #3}
\def\PRL#1#2#3{Phys. Rev. Lett. \underline{#1} (19#2) #3}
\def\Re{\mathop{\rm Re}}
\def\secder#1#2#3{{\partial^2 #1\over\partial #2 \partial #3}}
\def\s2w{\sin^2 \theta_W}
\def\Tr{\mathop{\rm Tr}}
\def\und{\underline}
\def\VEV#1{\left\langle #1\right\rangle} \let\vev\VEV
\def\mbf#1{\hbox{\boldmath $#1$}}
\relax
\def\dalpha{{\dot\alpha}}
\def\dbeta{{\dot\beta}}
\def\drho{{\dot\rho}}
\def\dsigma{{\dot\sigma}}
\def\crbig{\\\noalign{\vspace {3mm}}}
\def\bigint{{\displaystyle\int}}
\def\S{\Sigma}
\def\G{\Gamma}
\def\L{{\cal L}}
\def\SG{S_{\Gamma}}
\def\Fint{{\bigint d^2\theta\,}}
\def\Fbarint{{\bigint d^2\ov\theta\,}}
\def\Dint{{\bigint d^2\theta d^2\ov\theta\,}}

\title{ \vspace*{-0.8cm}
\begin{flushright}
\normalsize{CERN-PH-TH/2007-163}\\ 
\end{flushright}
\vspace{1cm}
\bf{Supersymmetric SU(5) GUT with Stabilized Moduli} \vspace*{-0.3cm}}

\begin{document}
\author{\bf\large{
Ignatios Antoniadis$^{1}$\footnote{On leave from CPHT (UMR du CNRS 7644), 
Ecole Polytechnique, F-91128 Palaiseau Cedex}
\footnote{Ignatios.Antoniadis@cern.ch}~,
Alok Kumar$^{2}$\footnote{kumar@iopb.res.in}~, 
Binata Panda$^{1,2}$\footnote{Binata.Panda@cern.ch, 
binata@iopb.res.in}\footnote{CERN Marie Curie fellow}}\\  
\\[-3mm]
\emph{\normalsize $^1$Department of Physics, CERN - Theory Division, 
CH-1211 Geneva 23, Switzerland }\\
\emph{\normalsize $^2$Institute of Physics, Bhubaneswar 751 005, 
India}\\
}

\date{\today}

\maketitle
\thispagestyle{empty}

\begin{abstract}
We construct a minimal example of a supersymmetric grand unified model 
in a toroidal compactification of type I string theory with magnetized 
$D9$-branes. All geometric moduli are stabilized in terms of the background 
internal magnetic fluxes which are of ``oblique" type 
(mutually non-commuting). The gauge symmetry is just $SU(5)$ and the 
gauge non-singlet chiral spectrum contains only three families of 
quarks and leptons transforming in the ${\bf 10} + {\bf\bar{5}}$ representations.
\end{abstract}
\date

\newpage
\section{Introduction}

Closed string moduli stabilization has been intensively studied during the  
last years for its implication towards a comprehensive understanding
of the superstring vacua\cite{GKP,KKLT}, 
as well as due to its significance in deriving definite low energy predictions
for particle models derived from string theory. Such 
stabilizations employ various supergravity\cite{GKP,KST}, non-perturbative\cite{KKLT} 
and string theory\cite{AM,AKM,Bianchi:2005yz} techniques to 
generate potentials for the moduli fields. However, very few examples are 
known so far of a complete stabilization of closed
string moduli in any specific model, while the known ones are too constrained
to accommodate interesting models from physical point
of view. Hence, there have been very few attempts to construct
a concrete model of particle physics even with partially 
stabilized moduli. Nevertheless, in view of the importance of 
the task at hand, we revisit the type I string 
constructions\cite{Angelantonj:2002ct,reviews} 
with moduli stabilizations\cite{AM,AKM,Bianchi:2005yz}, 
to explore the possibility of incorporating particle 
physics models, such as the Standard Model or GUT models based on 
grand unified groups, in  such a framework.

A new calculable method of moduli stabilization was recently proposed, 
using constant internal magnetic fields in four-dimensional (4d) 
type I string compactifications\cite{AM,AKM}. In the generic Calabi-Yau case, 
this method can stabilize mainly K\"ahler moduli\cite{Blumenhagen:2003vr,AM} and is thus 
complementary to 3-form closed string fluxes that stabilize the 
complex structure and the dilaton\cite{KST}. On the other hand, it can also 
be used in simple toroidal compactifications, stabilizing all geometric 
moduli in a supersymmetric vacuum using only magnetized $D9$-branes that 
have an exact perturbative string description\cite{Fradkin:1985qd, Bachas:1995ik}. 
Ramond-Ramond (RR) tadpole 
cancellation requires then some charged scalar fields from the branes to 
acquire non-vanishing vacuum expectation values (VEVs), breaking partly 
the gauge symmetry in order to preserve supersymmetry\cite{AKM}. Alternatively, one 
can break supersymmetry by D-terms and fix the dilaton at weak string 
coupling, by going ``slightly" off-criticality and thus generating a 
tree-level bulk dilaton potential\cite{ADM}.

There are two main ingredients for this approach of moduli stabilization\cite{AM,AKM}: 
(1) A set of nine magnetized $D9$-branes is needed to stabilize all 
36 moduli of the torus $T^6$ by the supersymmetry 
conditions\cite{MMMS,Angelantonj:2000hi}. Moreover, 
at least six of them must have oblique fluxes given by mutually 
non-commuting matrices, in order to fix all off-diagonal components 
of the metric. On the other hand, all nine $U(1)$ brane factors become 
massive by absorbing the RR partners of the K\"ahler class moduli\cite{Angelantonj:2000hi}. 
(2) Some extra branes are needed to satisfy the RR tadpole cancellation 
conditions, 
with non-trivial charged scalar 
VEVs turned on in order to maintain supersymmetry.

In this work, we apply the above method to construct phenomenologically 
interesting models. In the minimal case, three stacks of branes are 
needed to embed locally the Standard Model (SM) gauge group and the 
quantum numbers of quarks and leptons in their intersections\cite{AS}. 
They give rise to the gauge group $U(3)\times U(2)\times U(1)$, with the hypercharge 
being a linear combination of the three $U(1)$'s. Three different models 
can then be obtained, one of which corresponds to an $SU(5)$ Grand 
Unified Theory (GUT) when $U(3)$ and $U(2)$ are coincident. Here, 
we focus precisely on this $U(5)\times U(1)$ model employing two 
magnetized $D9$-brane stacks. Open strings stretched in the intersection 
of $U(5)$ with its orientifold image give rise to 3 chiral generations 
in the antisymmetric representation $\bf 10$ of $SU(5)$, while the 
intersection of $U(5)$ with the orientifold 
image of $U(1)$ gives 3 chiral states transforming 
as $\bf\bar 5$. Finally, the intersection of $U(5)$ with the 
$U(1)$ is non chiral, giving rise to Higgs pairs ${\bf 5}+{\bf\bar{5}}$.

In order to obtain an odd number (3) of fermion generations, a 
NS-NS (Neveu-Schwarz) 2-form $B$-field background\cite{Bquant} 
must be turned on\cite{RLBD}. 
This requires the generalization of the minimal set of branes with 
oblique magnetic fluxes that generate only diagonal 5-brane tadpoles 
on the three orthogonal tori of $T^6=\prod_{i=1}^3 T_i^2$. We find 
indeed a set of eight such ``oblique" branes which combined with $U(5)$ 
can fix all geometric moduli by the supersymmetry conditions. The metric 
is fixed in a diagonal form, depending on six radii given in terms 
of the magnetic fluxes. At the same time, all nine corresponding $U(1)$'s 
become massive yielding an $SU(5)\times U(1)$ gauge symmetry. This 
$U(1)$ factor cannot be made supersymmetric without the presence of 
charged scalar VEVs. Moreover, two extra branes are needed for RR 
tadpole cancellation, which also require non-vanishing VEVs to be made 
supersymmetric. As a result, all extra $U(1)$'s are broken and the 
only leftover gauge symmetry is an $SU(5)$ GUT. Furthermore, the 
intersections of the $U(5)$ stack with any additional brane used for 
moduli stabilization are non-chiral, yielding the three families of 
quarks and leptons in the ${\bf 10} + {\bf\bar{5}}$ representations as the only 
chiral spectrum of the model (gauge non-singlet).


To elaborate further,
the model is 
described by twelve stacks of branes, namely 
$U_5, U_1$, $O_1\dots , O_8$, $A$, and $B$. The $SU(5)$ gauge group 
arises from the open string states of stack-$U_5$ containing five 
magnetized branes. The remaining eleven stacks 
contain only a single magnetized brane. Also, the stack-$U_5$ containing 
the GUT gauge sector, contributes to the GUT particle spectrum through
open string states which either start and end on itself\footnote{For
simplicity, we do not distinguish a brane stack with its orientifold image,
unless is explicitly stated.} or on the stack-$U_1$, 
having only a single brane and therefore contributing an 
extra $U(1)$. For this reason we will also refer to these stacks 
as $U_5$ and $U_1$ stacks.

The matter sector of the  $SU(5)$ GUT is specified by 3 generations 
of fermions in  the group representations $\bf{\bar{5}}$ and 
$\bf{10}$ of $SU(5)$, both of left-handed helicity. 
In the magnetized branes construction, 
the $\bf{10}$ dimensional (antisymmetric) representation
of left-handed fermions:
\begin{eqnarray}
\bf{10} \equiv \begin{pmatrix} 0 & u_3^c & u_2^c & u_1 & d_1\cr
     & 0 & u_1^c & u_2 & d_2 \cr
& & 0 & u_3 & d_3 \cr
& & & 0 & e^+ \cr
& & & & 0 \end{pmatrix}_L
\label{ten}
\end{eqnarray}
arises from the doubly charged open string states starting on the 
stack-$U_5$ and ending at its orientifold image: $U_5^*$ and vice verse.
They transform as $\bf{10}_{(2,0)}$ of $SU(5)\times U(1)\times 
U(1)$, where the first $U(1)$ refers to stack-$U_5$ and the second one
to stack-$U_1$, while the subscript denotes the corresponding $U(1)$ charges. 
The $\bf{\bar{5}}$ of $SU(5)$ containing left-handed 
chiral fermions, or alternatively the $\bf{5}$ with right-handed fermions:
\begin{eqnarray}
\bf{5} \equiv \begin{pmatrix} d_1\cr d_2 \cr d_3 \cr e^+ \cr
\nu^c \end{pmatrix}_R
\label{5}
\end{eqnarray}
are identified as states of open strings starting from 
stack-$U_5$ (with five magnetized branes) and ending on stack-$U_1^*$ 
(i.e. the orientifold image of stack-$U_1$) and vice verse. 
The magnetic fluxes along the various branes 
are constrained by the fact that the chiral fermion spectrum, 
mentioned above, of the $SU(5)$ GUT should arise from these two sectors only. 

Our aim, in this paper, is to give a supersymmetric construction which 
incorporates the above features of $SU(5)$ GUT while stabilizing all the 
K\"ahler and complex structure moduli. More precisely, for fluxes to be
supersymmetric, one demands that their holomorphic $(2,0)$ part vanishes.
This condition  then leads to  
complex structure moduli stabilization\cite{AM}. 
In our case we show that, for the fluxes we turn on, 
the complex structure $\tau$ of $T^6$ is fixed to 
\begin{equation}
	\tau = i\id_3,
\label{complex}
\end{equation}
with $\id_3$ being the $3\times 3$ identity matrix.

In this paper, we make use of the conventions given in Appendix A
of Ref.\cite{AKM}, for the parametrization of the torus $T^6$, as well as
for the general definitions of the K\"ahler and complex
structure moduli. In particular, the coordinates of three
factorized tori: $(T^2)^3 \in T^6$ are given by $x_i, y_i$
$i=1,2,3$ with a volume normalization:
\be
\int dx_1\wedge dy_1\wedge dx_2\wedge dy_2\wedge dx_3
\wedge dy_3 =1.
\label{ft1}
\ee

For K\"ahler moduli stabilization, we make use of the mechanism based 
on the magnetized $D$-branes supersymmetry conditions 
as discussed in \cite{MMMS,AM,AKM}. Physically this corresponds 
to the requirement of vanishing of the potential which is generated 
for the moduli fields from the  Fayet-Iliopoulos (FI) D-terms associated with the
various branes. Even in this simplified scenario, 
the mammothness of the exercise is realized by noting that every 
magnetic flux that is introduced along any brane also 
induces charges corresponding to lower dimensional branes, 
giving rise to new tadpoles that need to be canceled. 
In particular, 
for the type I string  that we are discussing, there are 
induced $D5$ tadpoles from fluxes along the magnetized $D9$ branes.
These fluxes, in turn, are forced to be non-zero not only in order to satisfy the
condition of zero net chirality among the $U_5$ and the extra brane 
stacks (except with the $U_1$),
but in order to implement the mechanism of complex structure and 
K\"ahler moduli stabilization, as well. Specifically, 
for stabilizing the non-diagonal components of the metric, 
one is forced to introduce `oblique' fluxes along the $D9$-branes, thus
generating `oblique' $D5$-brane tadpoles, and all these need to be
canceled.

However, as mentioned earlier, we
are able to find eight brane stacks $O_1,\dots,O_8$, 
with different oblique fluxes, such that the combined net induced 
$D5$-brane charge lies only along the three diagonal directions $[x_i,y_i]$.
The holomorphicity conditions of fluxes, namely the vanishing
of  field strengths with
purely holomorphic indices, for these
brane stacks stabilizes the complex structure moduli to the
value (\ref{complex}).
These fluxes also introduce D-term potential for the 
K\"ahler moduli. Once the complex structure is fixed as in (\ref{complex}), 
the fluxes in the nine stacks $U_5, O_1,\dots,O_8$ generate potential 
in such a a way that all the nine K\"ahler moduli, 
$J_{i\bar{j}}$, ($i,j =1, 2, 3$) are completely fixed
by the D-flatness conditions, imposing the vanishing of the FI terms.
The residual diagonal tadpoles of the branes in the stacks
$U_5$, $U_1$, $O_1,\dots,O_8$ are then canceled by introducing the last two brane
stacks $A$ and $B$. 
D-flatness conditions for the brane stacks $U_1$, 
$A$ and $B$ are also satisfied, 
provided  some 
VEVs of charged scalars living on these branes are turned on to cancel the corresponding
FI parameters.
Magnetized $D$-branes provide exact CFT (conformal field theory) 
construction of the GUT model. However, 
in the presence of the these non-vanishing scalar VEVs,
exact CFT description is lost. The validity of 
the approximation then requires these VEVs 
to be smaller than unity in string units, a condition which is met in our case.
We explicitly determine the charged scalar VEVs and verify 
that they all take values $v^a << 1$.  
Our model therefore corresponds to the Higgsing of a magnetized $D9$-brane 
model to be made supersymmetric through the VEVs of certain charged scalar 
fields on the intersections of the branes $U_1, A$ and $B$.

At this point we would like to point out that, our strategy in this paper
is to start with a suitable ansatz for both the complex structure 
(\ref{complex}) and K\"ahler moduli leading to diagonal internal metric. 
Using this ansatz, we then determine fluxes along
the branes satisfying all the constraints we elaborated upon earlier.
We then use the flux solutions, to show explicitly
that the moduli are indeed 
completely fixed, consistent with our ansatz.

The rest of the paper is organized as follows. In 
Section \ref{preliminaries}, we give the necessary
constraints needed for building the model. This includes the 
discussion on moduli stabilization in subsection \ref{stabilization},
the tadpole constraints in subsection \ref{tadpoles} and the fermion
degeneracies in subsection \ref{spectrum}. 
Since a crucial step in a three generation model building is the 
introduction of a NS-NS (Neveu-Schwarz) $B$-field background 
without which only even generation models
can be built, the effect of non-zero $B$ on the chirality and
tadpoles is summarized in subsection \ref{constantnsnsb}.
In Section \ref{3generation}, we obtain general solutions
for fluxes along magnetized $D9$-branes 
satisfying the constraints of Section \ref{preliminaries}.
Moduli stabilization is
discussed in Section \ref{moduli-stabilization}. 
In Section \ref{fiparameter}, the 
VEVs of charged scalars on the stacks $U_1$, $A$ and $B$ are determined.
Our conclusions are presented in Section \ref{conclusions}. 
In Appendix \ref{Appendix-A}, the
fluxes along branes are written explicitly for the stacks 
$O_1,\dots ,O_8$ and the associated $D5$-brane tadpoles 
are given. The absence of chiral fermions is also shown from these sectors.
In Appendix \ref{Appendix-B}, complex structure stabilization is shown 
explicitly using the fluxes given in Appendix \ref{Appendix-A}. Finally, the K\"ahler moduli 
stabilization is shown in Appendix \ref{Appendix-C} (as well as in 
Section \ref{moduli-stabilization}).

\section{Preliminaries}\label{preliminaries}

We now briefly review the string construction using magnetized branes,
and in particular the chiral spectrum that follows for such stacks of branes
due to the presence of magnetic fluxes. 

\subsection{Fluxes and windings}\label{generalsetup}

We first briefly describe the construction based 
on $D$-branes with magnetic fluxes
in type I string theory, or equivalently type IIB with orientifold $O9$-planes
and magnetized $D9$-branes, in a $ T^6$ compactification. 
Later on, in subsection \ref{constantnsnsb}, we study the introduction
of constant NS-NS $B$-field background in this setup.

The stacks of $D9$-branes are characterized by three independent sets of 
data: (a) their multiplicities $N_a$, (b) winding 
matrices $W_{I}^{\hat{I},\, a}$ and 
(c) 1st Chern numbers $m^a_{\hat{I} \hat{J}}$ of the $U(1)$ background 
on their world-volume $\Sigma^a$,  $a=1,\dots,K$.  In our case, as 
already stated earlier, we have $K=12$ stacks. Also, $I, \hat{I}$ run over the
target space and world-volume indices, respectively. 
These parameters are described below:

(a) Multiplicities:  The first 
quantity $N_a$ describes the rank of the the 
unitary gauge group  $U(N_a)$ on each $D9$ stack. 

(b) Winding Matrices: The second set of parameters $W_{I}^{\hat{I},\, a}$ is the 
covering of the world-volume of each stack of $D9$-branes on the ambient 
space. They are characterized by the wrapping numbers of the branes around 
the different 1-cycles of the torus, which are encoded in the covering matrices 
$W^{\hat{I},\; a}_{\; I}$ defined as
\be
\label{md9:W}
W^{\hat{I}}_{\; J} = {\partial \xi^{\hat{I}} 
\over \partial X^J} \quad \quad {\rm for} \;\;  {\hat{I}}, J=0,\dots ,9\, ,
\ee
where the coordinates on the world-volume are denoted by $\xi^{\hat{I}}$, 
while the coordinates of the space-time ${\cal M}_{10}$ are $X^I$. Since 
space-time is assumed to be a direct product of a four-dimensional 
Minkowski manifold with a six-dimensional torus,  the 
covering matrix is of the form:
\be
\label{md9:W2}
W_{J}^{\hat{I},\, a} = 
\left(
\begin{array}{cc}
\delta_{\mu}^{\hat{\mu}} & 0 \\
0 & W_{\alpha}^{\hat\alpha\, , a}
\end{array}
\right)
\quad \quad {\rm for} \;\; \mu, \hat{\mu} = 0,\dots , 3 \;\; {\rm and } \;\; 
\alpha , \hat{\alpha} = 1,\dots , 6\, , 
\ee
with the upper block corresponding to the covering of $\Sigma_4^a$ on 
the four-dimensional 
space-time ${\cal M}_4$. Since these are assumed to be identical, 
the associated 
covering map $W^{\hat{\mu}}_\mu$ is the identity, 
$W^{\hat{\mu}}_\mu = \delta^{\hat{\mu}}_\mu$. The entries of the lower block, 
on the other hand, describe the wrapping numbers of the $D9$-branes around 
the different 1-cycles of the torus $T^6$ which are therefore restricted 
to be integers $W^{\hat{\alpha}}_\alpha \in \mathbb{Z}$,  $\forall\;  \alpha, 
\hat{\alpha} = 1,\dots, 6$ \cite{Bianchi:2005yz}. 

For simplicity, in the examples we consider here,
the winding matrix $W^{\hat{\alpha}}_\alpha$ in the internal directions  
is also chosen to be a six-dimensional diagonal matrix, implying 
an embedding such that the 
six compact $D9$ world-volume coordinates 
are identified with those of the internal target space 
$T^6$, up to a winding multiplicity factor $n^a_{\alpha}$, for a brane stack-$a$:
\be
n^a_{\alpha} \equiv W_{\alpha}^{\hat{\alpha}, a}.
\label{diag-winding}
\ee
We will also use the notation
 \be
\hat{n}^a_1 \equiv n_1^a n_2^a,\;\;
\hat{n}^a_2 \equiv n_3^a n_4^a,\;\;
\hat{n}^a_3 \equiv n_5^a n_6^a,\;\;\,\,({\rm no\,\, sum\,\, on\,\, a})
\label{hat-n}
\ee
to define the diagonal wrapping of the $D9$'s on the three 
orthogonal $T^2$'s inside $T^6$, given by:
\begin{eqnarray}
x^i\equiv X^{\alpha},\,\,\,\alpha = 1, 3, 5\, ;\,\,\,\,
y^i \equiv X^{\alpha},\,\,\,\alpha =2, 4, 6\, ,
\label{xiyi}
\end{eqnarray}
with periodicities: $x^i = x^i + 1$, $y^i\equiv y^i + 1$:
\be
\mathbb{T}^{6} = {\mbox{\huge $\otimes$}}_{i=1}^3{\mathbb{T}_{i}^2}\, ,
\label{factor}
\ee
and coordinates of the orthogonal 2-tori
($T^2_i$) being ($x^i, y^i$) for $i = 1, 2, 3$.

For further simplification, in our example, we will choose for all stacks trivial windings:
\be
n_{\alpha}^{a} \equiv 
W_{\alpha}^{\hat{\alpha}, a} = 1,\,\,{\rm for}\,\,\alpha =1,..,6,
\;\;a = U_5,\,U_1,\,O_1\cdots O_8,\,A,\,B.
\label{n_a's}
\ee
However in this section, in order to describe the formalism, we keep still
general winding matrices $W_{\alpha}^{\hat{\alpha}, a}$.

(c) First Chern numbers: The
parameters $m^a_{\hat{I} \hat{J}}$ of the brane data given above 
are the 1st Chern numbers of the $U(1)\subset U(N_a)$ background on 
the world-volume of the 
$D9$-branes. For each stack 
$U(N_a)=U(1)_a\times SU(N_a)$, the $U(1)_a$ has a
constant field strength 
on the covering of the internal space. These are subject to the Dirac 
quantization condition which implies that all internal magnetic
fluxes $F^a_{\hat{\alpha}\hat{\beta}}$, 
on the world-volume of each stack of $D9$-branes, 
are integrally quantized. 

Explicitly, the world-volume fluxes $F^a_{\hat{\alpha}\hat{\beta}}$ and the 
corresponding target space induced fluxes  $p^a_{\alpha\beta}$
are quantized as
\be
\label{stab:gen:F}
\left\{
\begin{array}{ccc}
 F^a_{\hat{\alpha}\hat{\beta}} = m^a_{\hat{\alpha}\hat{\beta}} \in \;  
\mathbb{Z}&\; \forall\, \hat{\alpha},\hat{\beta} = 1,\dots, 6&
\\
&&\quad \quad \forall a = 1,\dots, K\, .
\\
p^a_{\alpha\beta} =  (W^{-1})_{\alpha}^{\hat{\alpha}, \; 
a} (W^{-1})_\beta^{\hat{\beta} , \; a} \,  
m^a_{\hat{\alpha}\hat{\beta}} \in \mathbb{Q}, & \; \forall\, \alpha,
\beta = 1,\dots, 6&
\end{array}
\right.
\ee 
For later use, when fluxes are turned on only along three 
factorized $T^2$'s of eq. (\ref{factor}), as will be the case for some of 
our brane stacks, we make use of the following convenient notation:
\be
\hat{m}^a_1 \equiv m_{12}^a \equiv m^a_{x^1y^1},\;\;\,  
\hat{m}^a_2 \equiv m_{34}^a\equiv m^a_{x^2y^2},\;\;\,
\hat{m}^a_3 \equiv m_{56}^a\equiv m^a_{x^3y^3}.
\label{hat-m}
\ee

The magnetized $D9$-branes couple only to the $U(1)$ flux associated 
with the gauge fields located on their own world-volume. In other 
words, the charges  of 
the endpoints $q_R$ and $q_L $ of the open strings stretched 
between the $i$-th and the 
$j$-th $D9$-brane can be written as $q_L  \equiv q_i$ and 
$q_R \equiv -q_j$, while the Cartan 
generator $h$ is given by
$h = {\rm diag}(h_1\id_{N_1}, \dots , h_N\id_{N_K})$, with 
$\id_{N_a}$ being the $N_a\times N_a$ identity matrix.
In addition, in type I string theory, the number of magnetized $D9$-branes 
must be doubled. Since 
the orientifold projection ${\cal O}=\Omega_p$ is defined by the 
world-sheet parity, 
it maps the field strength $F_a =  dA_a$ of the $U(1)_a$ gauge potential 
$A_a$ to its opposite, 
${\cal O}: F_a \rightarrow -F_a$. Therefore, the magnetized $D9$-branes are 
not an invariant  configuration and for each stack a mirror stack 
must be added with opposite flux on its world-volume. 

\subsection{Stabilization}\label{stabilization}
We now write down the supersymmetry conditions for magnetized 
$D9$-branes in the context of type I toroidal 
compactifications and discuss the stabilization of complex structure 
and K\"ahler class moduli using such conditions.

The geometric  moduli of $T^6$ decompose in a complex structure 
variation which is parametrized by the matrix $\tau_{ij}$  
entering in the definition of the complex coordinates 
\be
z_i = x_i + \tau_{ij}y^j\, ,
\label{complexbasis}
\ee
and in the K\"ahler variation of the mixed part of the metric  described by the 
real $(1,1)$-form $J = i\delta g_{i\bar{j}} dz^i \wedge d\bar{z}^j$. 
The supersymmetry conditions then read\cite{AM,AKM}:
\be
F_{(2,0)}^a = 0\ \ ; \quad {\cal F}_a \wedge {\cal F}_a \wedge {\cal F}_a 
= {\cal F}_a \wedge J \wedge J\ \ ; \quad {\rm det}W_a 
\left(J \wedge J \wedge J - {\cal F}_a \wedge {\cal F}_a \wedge J\right)>
0 
\, ,
\label{susy}
\ee
for each $a = 1,\dots, K$. The complexified fluxes can be written as
\bea
F^a_{(2,0)} & = & {(\tau-\bar{\tau})^{-1}}^T \! \left[ \tau^{T} p^{a}_{xx} 
\tau - \tau^{T}{p^{a}_{xy}} - p^{a}_{yx}\tau + 
p^{a}_{yy}\right] (\tau-\bar{\tau})^{-1}, \label{stab:gen:purelyholo2}\\
F^a_{(1,1)} & = & {(\tau-\bar{\tau})^{-1}}^T\! \left[ -\tau^{T} 
p^{a}_{xx}\bar{\tau} + \tau^{T}{p^{a}_{xy}} + p^{a}_{yx}\bar{\tau} - 
p^{a}_{yy}\right] (\tau-\bar{\tau})^{-1},
\label{susy-kahler0}
\eea
where the matrices $(p^{a}_{x^ix^j})$, $(p^{a}_{x^iy^j})$ and 
$(p^{a}_{y^iy^j})$
are the quantized field strengths in target space, given in eq.
(\ref{stab:gen:F}). For our choice (\ref{n_a's}), they coincide with the
Chern numbers $m^a$ along the corresponding cycles. The field strengths
 $F^a_{(2,0)}$ and $F^a_{(1,1)}$ are $3\times 3$
matrices that correspond to the upper half of the matrix 
$\mathcal{F}^a$:
\be
\mathcal{F}^a\equiv -(2\pi)^2 i\alpha' \left(
\begin{array}{cc}
F^a_{(2,0)} & F^a_{(1,1)}\\
-{F^a}^\dagger_{(1,1)} & {{F^a}^*}_{(2,0)}\\
\end{array}
\right)\, ,
\label{matrixF}
\ee
which is the total field strength in the 
cohomology basis $e_{i\bar{j}} = i dz^i \wedge d\bar{z}^j$\cite{AM,AKM}. 

The first set of conditions of eq.~(\ref{susy}) states that the 
purely holomorphic 
flux vanishes. For given flux quanta and winding numbers, this matrix equation 
restricts the complex structure $\tau$. Using 
eq.~(\ref{stab:gen:purelyholo2}), 
the supersymmetry conditions for each stack can first 
be seen as a restriction on the 
parameters of the complex structure matrix elements $\tau$:
\be
F_{(2,0)}^a = 0\qquad  \rightarrow \qquad \tau^{T} p^{a}_{xx} \tau - 
\tau^{T}{p^{a}_{xy}} - p^{a}_{yx}\tau + 
p^{a}_{yy} = 0\, ,
\label{stab:gen:M20_condition}
\ee
giving rise to at most six complex equations for each brane stack $a$.

The second set of conditions  of eq.~(\ref{susy}) gives rise to a real 
equation and restricts the K\"ahler moduli. This can be understood as a 
D-flatness condition. In the four-dimensional effective action, the 
magnetic fluxes give rise to  topological couplings for the 
different axions of 
the compactified field theory. These arise from the dimensional 
reduction of the 
Wess Zumino action. In addition to the topological coupling, 
the ${\cal N}=1$ 
supersymmetric action yields a Fayet-Iliopoulos (FI) term of the form: 
\be
{\xi_a\over g_a^2} = {1\over (4\pi^2 \alpha')^3}
\int_{T^6}\big( {\cal F}_a \wedge {\cal F}_a\wedge {\cal F}_a -
{\cal F}_a\wedge J \wedge J\big)\, . 
\label{xioverg2}
\ee
The D-flatness condition in the  absence of charged scalars requires then 
that $<{\rm D}_a> = \xi_a = 0$, which is equivalent to the second 
equation of eq.~(\ref{susy}). Finally, the last inequality  
in eq.~(\ref{susy}) may also be understood from a four-dimensional 
viewpoint  as the positivity of the $U(1)_a$ gauge coupling $ g_a^2$, since 
its  expression  in terms of the fluxes and moduli reads
\be
{1\over g_a^2} = {1\over (4\pi^2 \alpha')^3} 
\int_{T^6}\big( J \wedge J\wedge J -{\cal F}_a\wedge {\cal F}_a \wedge J\big)\, .
\label{gaugecoupling}
\ee

The above supersymmetry conditions, get modified
in the presence of VEVs for scalars charged under 
the $U(1)$ gauge groups of the branes. The D-flatness condition, 
in the low energy field theory approximation, then reads: 
\begin{equation}
{\rm D}_a = - \left( \sum_\phi q^\phi_a |\phi|^2G_\phi + M_s^2\xi_a \right) =0 \, ,
\label{dterm}
\end{equation}
where $M_s=\alpha'^{-1/2}$ is the string scale\footnote{When mass scales are absent, string units are implicit throughout the paper.}, 
and the sum is extended over all
scalars $\phi$ charged under the $a$-th $U(1)_a$ with charge $q_a^{\phi}$
and metric $G_\phi$. 
Such scalars arise in the compactification of magnetized $D9$-branes in 
type I string theory, for instance from the NS sector of open 
strings stretched 
between the $a$-th brane and its image $a^\star$, or between the
stack-$a$ and 
another stack-$b$ or its image $b^*$. When one of these scalars acquire 
a non-vanishing VEV $\langle|\phi|\rangle^2 = v_\phi^2$,  the calibration condition of 
eq.~(\ref{susy}) is modified to:
\bea
q_a v_a^2 \int_{T^6}\big( J \wedge J\wedge J -
{\cal F}_a\wedge {\cal F}_a \wedge J\big)\!\! &=&\!\! -
\frac {M_s^2}{G}{\int_{T^6}\big( {\cal F}_a \wedge {\cal F}_a\wedge {\cal F}_a -
{\cal F}_a\wedge J \wedge J\big)}
\label{stab:gen:cond_kahlerX} 
\\ [0.3cm] 
{\det W}_a\left(J \wedge J \wedge J - 
{\cal F}_a \wedge {\cal F}_a \wedge J \right)\!\! &>&\!\! 0 \quad , \, 
\quad \forall a=1,\dots, K \, .
\label{stab:gen:cond_pos}
\eea
Note that our computation is valid for small values
of $v_a$ (in string units), since the inclusion of the charged scalars in the
D-term is in principle valid perturbatively. 

Actually, the fields appearing in (\ref{dterm}) are not canonically normalized since
the metric $G_\phi$ appears explicitly also in their kinetic terms. Thus, the physical
VEV is $v_\phi\sqrt{G_\phi}$. However, to estimate the validity of the perturbative approach,
it is more appropriate to keep $v_\phi$ instead of $v_\phi \sqrt {G_\phi}$. The reason is that
the next to leading correction to the D-term involves a quartic term of the type $|\phi|^4$,
proportional to a new coefficient $\cal K$, and the condition of validity of perturbation
theory is ${\cal K}v_\phi^2/G_\phi<<1$. A rough estimate is then obtained by approximating 
${\cal K}\sim G_\phi$, which gives our condition.

The metric $G_\phi$ of the scalars
living on the brane has been computed explicitly for the case of 
diagonal fluxes\cite{0404134}. In this special case, the fluxes are
denoted by three angles $\theta_i^a$, ($i=1,2,3$).\footnote{See examples in 
Appendix A for the precise map between $p_{x^iy^i}$ and 
$(F_{(1, 1)})_{z^i\bar{z}^i}$.} Then suppressing index-$a$, we have:
\be
	{\rm tan}\, \pi \theta_i = \frac{p_{x^iy^i}}{J_i} \equiv 
		\frac{{(F_{(1, 1)})}_{z^i\bar{z}^i}}{J_i} ,
\label{theta}
\ee
and 
\be
	G = e^{\gamma_E (\theta_1 + \theta_2 + \theta_3)}\times
	\sqrt{\frac{\Gamma(\theta_1)\Gamma(\theta_2)\Gamma(\theta_3)}
	{\Gamma(1 - \theta_1)\Gamma(1 - \theta_2)\Gamma(1 - \theta_3)}},
\label{metric}
\ee
with $\gamma_E$ being the Euler constant.
The above results will be applied in section \ref{fiparameter} to find out the FI 
parameters and charged scalar VEVs along three of the twelve
brane stacks: $U_1$, $A$ and $B$. The other nine stacks, 
$U_5$, $O_1,\dots , O_8$, stabilizing
all the geometric moduli, will satisfy the calibration 
condition 
$\xi^a = 0$ in the absence of open string scalar VEVs.
Moreover, the RR moduli that appear in the same chiral multiplets as the geometric
K\"ahler moduli, become Goldstone modes which get absorbed by the $U(1)$ 
gauge bosons\cite{AM} corresponding to each of the D-terms that stabilize
the relevant geometric moduli.

\subsection{Tadpoles}\label{tadpoles}

In toroidal compactifications of type I string 
theory, the magnetized $D9$-branes induce 5-brane 
charges as well, while the 
3-brane and 7-brane charges automatically vanish due to the presence of mirror 
branes with opposite flux. For general magnetic fluxes, RR
tadpole conditions can be written in 
terms of the Chern numbers and winding matrix \cite{Bianchi:2005yz,AKM}
as:
\bea
16 &=& \sum_{a=1}^K \; N_a\;  {\rm det}W_a
\equiv \sum_{a=1}^K \;  Q^{9,\, a},
\label{tad9}
\\
0  &=& \sum_{a=1}^K\; N_a \; {\rm det}W_a \; 
{\cal Q}^{a,\,\,\alpha\beta} \equiv \sum_{a=1}^K\; 
{Q}^{5,\, a}_{\alpha\beta}, \qquad \forall \alpha,\beta=1,\dots,6
\label{tad5}
\eea
where
\be
{\cal Q}^{a,\,\,\alpha\beta} = \epsilon^{\alpha\beta\delta\gamma\sigma\tau} 
p^a_{\delta\gamma}p^a_{\sigma\tau} \, .\nonumber
\ee
The l.h.s. of eq.~(\ref{tad9}) arises from the contribution of the 
$O9$-plane. On the other hand, in toroidal compactifications there are no 
$O5$-planes and thus  the l.h.s. of eq.~(\ref{tad5}) vanishes. 

For our choice of windings (\ref{n_a's}), $W_i^{\hat{i}} = 1$, the $D9$ tadpole 
contribution from a given stack-a of branes is simply equal to the 
number of branes, $N_a$. The $D5$ tadpole expression 
also takes a simple form for the fluxes satisfying the 
$F^a_{(2, 0)} = 0$ condition (\ref{susy}). The fluxes 
are then represented by three-dimensional Hermitian matrices 
$(F^a_{(1, 1)})$ which appeared in eq. (\ref{matrixF}) and
the $D5$ tadpoles ${\cal Q}^{5,\, a}_{i\bar{j}}$ are the Cofactors of the 
${i\bar{j}}$ matrix elements $(F^a_{(1, 1)})_{i\bar{j}}$. 
Fluxes and tadpoles
in such a form  are given in Appendix A.

\subsection{Spectrum}\label{spectrum}

The gauge sector of the spectrum follows from the open string states
corresponding to strings starting and ending on the same brane stack.
The gauge symmetry group is given by a  product of unitary groups 
$\otimes_a U(N_a)$, 
upon identification of the associated open strings  attached 
on a given stack  with the ones attached on the mirror 
(under the orientifold transformation) stack. 
In addition to these vector bosons, the massless spectrum contains adjoint scalars 
and fermions forming ${\cal N}=4$, $d=4$ supermultiplets.

In the matter sector, the massless spectrum is obtained from the following 
open string states\cite{bi,Angelantonj:2000hi}:
\begin{enumerate}
\item Open strings stretched between 
the $a$-th and $b$-th stack give rise to chiral spinors in the 
bifundamental representation 
$(N_a,\bar{N}_b)$ of $U(N_a)\times U(N_b)$. Their multiplicity $I_{ab}$ is given 
by \cite{Bianchi:2005yz}:
\be
I_{ab} = 
{ {\rm det}W_a{\rm det}W_b \over (2\pi)^3} \int_{T^6}
\left( q_a F^a_{(1, 1)} +q_b F^b_{(1, 1)}\right)^3\, ,
\label{intersection}
\ee
where $F^{a}_{(1, 1)}$ (given in eqs. (\ref{susy-kahler0}) and (\ref{matrixF}))
is the pullback of the integrally quantized world-volume flux 
$m^a_{\hat{\alpha} \hat{\beta}}$ on the target torus in the complex basis (\ref{complexbasis}), 
and $q_a$ is the corresponding $U(1)_a$ charge; in our case $q_a=+1$ $(-1)$ for the 
fundamental (anti-fundamental representation). The transformation under the gauge 
group and their multiplicities are thus determined in terms of the data 
$(N_a, W_{I}^{\hat{I},\, a},m_{\hat{I} \hat{J}})$. 

For factorized toroidal compactifications $(T^2)^3$ (\ref{factor})
with only diagonal fluxes $p_{x^iy^i}$ $(i=1, 2, 3)$, the multiplicities of 
chiral fermions, arising from strings starting from stack $a$ and 
ending at $b$ or vice versa,
take the simple form (using notations of eqs.~(\ref{hat-n}) and (\ref{hat-m})):
\be
(N_a,{\overline N}_b)  :  I_{a b} = 
\prod_i (\hat{m}_i^a \hat{n}_i^b - \hat{n}_i^a 
\hat{m}_i^b),\nonumber\\
\ee
\be
(N_a, N_b)  :  I_{ab^*} = \prod_i (\hat{m}_i^a \hat{n}_i^b + \hat{n}_i^a
\hat{m}_i^b)\, .
\label{scmult}
\ee
where $i$ is the label of the $i$-th two-tori $T^2_i$, and 
the integers $\hat{m}_i^a,
\hat{n}_i^a$ enter in the multiplicity expressions through 
the magnetic field as in eq. (\ref{stab:gen:F}).

In the model that we construct, however, 
we need stacks with fluxes which contain both diagonal and 
oblique flux components, for the purpose of complete K\"ahler 
and complex structure moduli stabilization.

\item Open strings stretched between the $a$-th brane and its 
mirror $a^\star$ give rise 
to massless modes associated to $I_{aa^\star}$ chiral fermions. 
These transform either 
in the antisymmetric or symmetric representation of $U(N_a)$. 
For factorized toroidal
compactifications $(T^2)^3$, the multiplicities of chiral fermions 
are given by;
\be
{\rm Antisymmetric}  : \quad {1\over 2}\left(\prod_i
2\hat{m}_i^a\right)\left(\prod_j \hat{n}_j^a+1\right),\nonumber \\
\ee
\be
{\rm
Symmetric}  :  \quad {1\over 2}\left(\prod_i 2\hat{m}_i^a\right)\left(\prod_j
\hat{n}_j^a-1\right).
\label{dcmult}
\ee
\end{enumerate}
In generic configurations, where supersymmetry is broken by 
the magnetic fluxes, the scalar partners of the massless chiral spinors
in twisted open string sectors ({\em i.e.} from non-trivial brane intersections)
are massive (or tachyonic).
Moreover, when a chiral index $I_{ab}$ vanishes, the corresponding intersection of stacks $a$ and $b$ is non-chiral. The multiplicity of the non-chiral spectrum is then determined by extracting the vanishing factor and calculating the corresponding chiral index in higher dimensions. This is done explicitly for our model below, in section \ref{non-chiral-spectrum}.

\subsection{Constant NS-NS B-field backround}\label{constantnsnsb}

In toroidal models with vanishing $B$-field, the 
net generation number of chiral fermions is in general even\cite{RLBD}.  
Thus, it is necessary to turn on a constant $B$-field background in order to obtain a
Standard Model like spectrum with three generations.  Due to the world-sheet parity
projection $\Omega$, the NS-NS two-index field $B_{\alpha \beta}$ is
projected out from the physical spectrum and constrained to take the discrete values
$0$ or $1/2$ (in string units) along a 2-cycle $(\alpha \beta)$ of $T^6$~\cite{Bquant}. 

For branes at angles,
$B_{\alpha \beta}=1/2$ changes the number of intersection 
points of the two branes. 
For the case of magnetized $D9$-branes, 
if $B$ is turned on only along the three diagonal 2-tori:
\be
B_{x^iy^i} \equiv b_i = \frac 12\, ,\,\,\, i=1, 2, 3, 
\label{diagonalB}
\ee 
the effect is accounted for by
introducing an effective world-volume magnetic flux
quantum, defined by 
${\tilde{\hat{m}}}^a_j=\hat{m}^a_j+\frac 12 \hat{n}^a_j$, while
the first Chern numbers along all other 2-cycles remain unchanged (and integral).
Thus, the modification can be summarized by 
\be
(\hat{m}^a_j,\hat{n}^a_j)\,\,\,\, {\rm for}\,\,\, b_{j}=0\,\,\,\,
\rightarrow   (\hat{m}^a_j+\frac 12 \hat{n}^a_j, \hat {n}^a_j)
\equiv ({\tilde{\hat{m}}}^a_j,\hat{n}^a_j)\, ,\,\,\,
{\rm for}\,\,  b_j = \frac 12, 
\label{b-transform}
\ee
along the particular 2-cycles where the NS-NS $B$-field is turned on.
This transformation also takes into account 
the changes in the fermion degeneracies given in 
eqs.~(\ref{scmult}) and (\ref{dcmult})
(as well as in (\ref{intersection-ab}), (\ref{intersection-ab'})
below), due to the presence of a non-zero $B$:
\be
(N_a,{\overline N}_b)  : I_{ab} = \prod_i (\tilde{\hat{m}}_i^a {\hat{n}}_i^b -
{\hat{n}}_i^a 
\tilde{\hat{m}}_i^b),\nonumber\\
\ee
\be
(N_a,N_b)  :  I_{ab^*} = \prod_i (\tilde{\hat{m}}_i^a {\hat{n}}_i^b
+{\hat{n}}_i^a
\tilde{\hat{m}}_i^b)\, ,
\label{scmult'}
\ee
\be
{\rm Antisymmetric}  :  I^A_{aa^*} = \quad {1\over 2}\left(\prod_i 2
\tilde{\hat{m}}_i^a \right)\left(\prod_j \hat{n}_j^a + 1\right),
\label{dcmult''}
\ee
\be
{\rm Symmetric}  :  I^S_{aa^*} = \quad {1\over 2}\left(\prod_i 2
\tilde{\hat{m}}_i^a\right)\left(\prod_j \hat{n}_j^a-1\right).
\label{dcmult'}
\ee
In addition, similar modifications take place
in the tadpole cancellation conditions, as well.
Note that for non trivial $B$, if $\hat{n}_i^a$ is odd $\tilde{\hat{m}}_i^a$ 
is half-integer, while if $\hat{n}_i^a$ is even $\tilde{\hat{m}}_i^a$ must be integer.

When restricting to the trivial windings of eq.~(\ref{n_a's}) that we use in this paper, 
$\hat{n}_i^a=1$, the degeneracy formula (\ref{intersection}) simplifies to:
\be
(N_a, {\overline N}_b)  : I_{ab} = \det \left({\tilde F}^a_{(1, 1)} - 
				{\tilde F}^b_{(1, 1)}\right)\, ,
\label{intersection-ab}
\ee
\be
(N_a, N_b)  : I_{ab^*} = \det \left({\tilde F}^a_{(1, 1)} + {\tilde F}^b_{(1, 1)}\right),\, 
\label{intersection-ab'}
\ee
where ${\tilde F}=F+B$ and
we have assumed the canonical volume normalization (\ref{ft1}) on $T^6$.
Similarly, the multiplicity of chiral antisymmetric representations is given by:
\be
{\rm Antisymmetric}  :\quad  I^A_{aa^*} = \prod_i \left( 2\tilde{\hat{m}}_i^a\right)\, ,
\label{intersection-aa*}
\ee
while there are no states in symmetric representations. Finally,
the tadpole cancellation conditions (\ref{tad9}) and (\ref{tad5}) become:
\be
\sum_{a=1}^K \; N_a = 16\quad ;\quad
\sum_{a=1}^K \; N_a\; {\rm Co}({\tilde F}^a_{(1, 1)})_{i\bar{j}} = 0 \qquad
\forall\, i,j=1,\dots,3\, .
\label{tad95}
\ee

\section{Constructing a three generation $SU(5)$ GUT model}\label{3generation}

In this section, we first present in subsection \ref{su5gut} 
the brane stacks $U_5$ and $U_1$, on which
the $SU(5)$ GUT, with three  generations of chiral fermions, lives.  
Then, in subsection \ref{nonchiral}, we write down the conditions 
which any extra stacks,
called $O_a$ have to satisfy, so that there are no net $SU(5)$ non-singlet 
chiral fermions corresponding to open strings of the type:
$U_5-O_a$ and $U_5-O_a^*$. In other words:
\be
	I_{U_5O_a} + I_{U_5O_a^{*}} = 0.
\label{zero-chirality}
\ee
In addition, we also write down, in subsection \ref{susy-constraint}, 
the condition that such
stacks are mutually supersymmetric with the stack $U_5$, without turning on 
any charged scalar VEVs on these branes.  
The solution of these
conditions giving eight branes $O_1,...,O_8$ is presented in 
subsections \ref{solutions} and \ref{additional}.
They are all supersymmetric, stabilize all K\"ahler
moduli (together with stack-$U_5$)
and cancel all tadpoles along the oblique 
directions, $x_ix_j$, $x_i y_j$, $y_i y_j$  for $i\neq j$.
Finally in subsection \ref{tadpolecancellation}, 
two more stacks A and B are found which cancel
the overall $D9$ and $D5$-brane tadpoles (together with the $U_1$ stack).

As stated earlier,
our strategy to find solutions for branes and fluxes is to first assume a 
canonical complex structure and K\"ahler moduli 
which have non-zero components only along the three factorized orthogonal 2-tori.
In other words, we look for solutions where K\"ahler moduli are 
eventually stabilized such that
\be
J_{i\bar{j}} = 0,\,\, i\neq j,\,\,\,(i,j=1,2,3).
\label{oblique-kahler}
\ee
By assuming the complex structure and K\"ahler moduli as in 
eqs. (\ref{complex}) and (\ref{oblique-kahler}), 
we then find fluxes needed to be turned on in order to 
cancel tadpoles. These fluxes are also
used in the stabilization equations, in 
section \ref{moduli-stabilization} and Appendices 
\ref{Appendix-B} and \ref{Appendix-C}, to show that 
moduli are indeed completely fixed in a way that the six-torus metric becomes diagonal.

\subsection{SU(5) GUT brane stacks}\label{su5gut}

We now present the two brane stacks $U_5$ and $U_1$ which give the particle
spectrum of $SU(5)$ GUT. For this purpose,
we consider diagonally magnetized $D9$-branes on a factorized six-dimensional internal torus
({\ref{factor}}), in the presence of a
NS-NS $B$-field turned on according to eq. (\ref{diagonalB}).
The stacks of $D9$-branes have multiplicities
$N_{U_5}$ = 5 and $N_{U_1}$ = 1,
so that an $SU(5)$ gauge group can be accommodated on the first one.
Next, we impose a constraint on the windings $\hat{n}_i^{U_5}$
(defined in eq.(\ref{hat-n}))
of this stack by demanding that chiral fermion multiplicities
in the symmetric representation of $SU(5)$ is zero. Then from  
eqs. (\ref{dcmult'}), 
we obtain the constraint: 
\be
\prod_j \hat{n}_j^{U_5} = 1.
\label{winding-constraint}
\ee
We solve eq. (\ref{winding-constraint})
by making the choice (\ref{n_a's}): $n_{\alpha}^{U_5} 
\equiv W_{\alpha}^{\hat{\alpha}, U_5} = 1$ 
for the stack $U_5$. This also implies $\hat{n}_i^{U_5} =1$ for $i=1, 2, 3$.
Moreover, since from (\ref{tad9}) the total $D9$-brane charge has
to be sixteen and higher winding numbers give larger contributions to the
$D9$ tadpole, the windings in all stacks will be 
restricted\footnote{detW is restricted to be positive definite in order
to avoid the presence of anti-branes.} 
to $n_i^a = 1$
so that a maximum number of brane stacks can be
accommodated (with $Q^9 = 16$), 
in view of fulfilling the task of stabilization.
 
Indeed, the stack $U_5$
already saturates five units of $D9$ charge while stabilizing only a 
single K\"ahler modulus. 
One more unit of $D9$ charge is saturated by the
$U_1$ stack, responsible for producing the chiral fermions in the representation
$\bf{\bar 5}$ of $SU(5)$ at its intersection with $U_5$.
Moreover, it cannot be made supersymmetric in the absence of charged scalar
VEVs, as we will see below. Thus,
stabilization of the eight remaining K\"ahler moduli, apart from the one 
stabilized by the $U_5$ stack, needs eight additional 
branes $O_1,\dots , O_8$, 
contributing at least that many units of $D9$ charge (when 
windings are all one). These leave only two units of $D9$ charge
yet to be saturated, which are also required to
cancel any $D5$-brane tadpoles 
generated by the ten stacks, $U_5, U_1$ and $O_1,\dots ,O_8$. 
We find that this 
is achieved by two stacks A and B, also of windings one, so that the
total $D9$ charge is $Q^9=16$ and all $D5$ tadpoles vanish $Q^5_{\alpha \beta} = 0$.  

Now, after having imposed the condition that symmetric doubly charged representations
of $SU(5)$ are absent, we find solutions for the
first Chern numbers and fluxes, so that the
the degeneracy of chiral fermions in the antisymmetric 
representation $\bf 10$ is equal to three. 
These multiplicities are given in eqs. (\ref{dcmult''}), (\ref{intersection-aa*}), and when 
applied to the stack $U_5$ 
give the constraint:
\be
 (2 \hat{m}^{U_5}_1 + 1)(2 \hat{m}^{U_5}_2 + 1)(2 \hat{m}^{U_5}_3 + 1)  = 3,
\label{dcmult-constraint}
\ee
with a solution:
\be
   \hat{m}_1^{U_5} = -2,\,\,\,\hat{m}_2^{U_5} = -1,\,\,\,
\hat{m}_3^{U_5} = 0 .
\label{soln-chern}
\ee
The corresponding flux components are:
\be
	p_{x^1y^1}^{U_5} = -\frac{3}{2},\,\,\,
	p_{x^2y^2}^{U_5} = -\frac{1}{2},\,\,\,
	p_{x^3y^3}^{U_5} = \frac{1}{2},
\label{flux-u5}
\ee
associated to the total (target space) flux matrix
\begin{eqnarray}
	{\tilde F}^{U_5}_{(1, 1)} = \begin{pmatrix}-\frac{3}{2} & & \cr
	& -\frac{1}{2} & \cr
	& & \frac{1}{2} \end{pmatrix}.
\label{fu5ibarj}
\end{eqnarray}
At this level, the choice of signs is arbitrary and is taken for convenience.

Next, we solve the condition for the presence of three generations
of chiral fermions 
transforming in $\bf{\bar 5}$ of $SU(5)$.
These come from singly charged open string states
starting from the $U_5$ stack and ending on the $U_1$ stack
or its image. In other words, we use the condition:
\be
	I_{U_5U_1} + I_{U_5U_1^*} = -3.
\label{i51+i51'}
\ee
To solve this condition for diagonal fluxes, one can use the formulae (\ref{scmult'}),
or alternatively eqs. (\ref{intersection-ab}) and (\ref{intersection-ab'}). 
In the presence of the NS-NS $B_{\alpha \beta}$-field
of our choice (\ref{diagonalB}), 
and using the fluxes along the $U_5$ stack
(\ref{flux-u5}) or (\ref{fu5ibarj}), the formulae take a form:
\be
(N_{U_5}, {\overline N}_{U_1})  :\quad  I_{U_5U_1} = 
(-\frac 32 - F^{U_1}_1)(-\frac 12 - F^{U_1}_2)(\frac 12 - F^{U_1}_3)\, ,
\label{intersection-Bfab'}
\ee
\be
(N_{U_5}, N_{U_1})  :\quad  I_{U_5U_1^*} = 
(-\frac 32 + F^1_{U_1})(-\frac 12 + F^{U_1}_2)
(\frac 12 + F^{U_1}_3)\, ,
\label{intersection-Bfab}
\ee
where we have used the notation $F_i^a \equiv ({\tilde F}^a_{(1, 1)})_{i\bar{i}}$ 
for a given stack-a. We will also demand that all components 
$F^{U_1}_1, F^{U_1}_2, F^{U_1}_3$ are half-integers, due to the shift in 1st Chern numbers 
$\hat{m}_i^{U_1}$ by half a unit,
in the presence of a non-zero NS-NS $B$-field along the three 
$T^2$'s (\ref{factor}). We then get a solution of eq. (\ref{i51+i51'}):
\be
	I_{U_5U_1} = 0,\,\,\, I_{U_5U_1^*} = -3,
\label{i51,i51'}
\ee
for flux components on the stack $U_1$:
\be
	F_1^{U_1} = -\frac{3}{2},\,\,\,F_2^{U_1} = \frac{3}{2},\,\,\,
	F_3^{U_1} = \frac{1}{2}.
\label{flux-u1}
\ee

One  can ask whether solutions other 
than (\ref{flux-u1}) are possible for the $U_1$ stack.
For instance, instead of the choice $(0,-3)$ of eq.~(\ref{i51,i51'}) for the 
intersections $U_5-U_1$ and $U_5-U_1^*$ subject to the condition
(\ref{i51+i51'}), one could try $(-3,0)$ or in general $(n,-n-3)$, for $n$
any integer. Note that $n$ (for $n>0$) or $-n-3$ (for $n<-3$) is the number
of electroweak Higgs pairs contained in ${\bf 5}+{\bf{\bar 5}}$ of $SU(5)$.
Thus, the cases $(-1,-2)$ and $(-2,-1)$ were excluded because of the
absence of higgses, but other cases such as $n=1$ or $n=-4$ (containing
one Higgs pair) are worth to explore. We leave these as exercises for the future.

The present results, including the quanta $(\hat{m_i}, \hat{n_i})$ 
for both $U_5$ and $U_1$ stacks, are summarized 
in Table~\ref{table-1}.
\begin{table} 
\vskip 0.4cm
\begin{center}
\begin{tabular}{|c|c|c|c|c|}
\hline
Stack no. & No. of & Windings  & Chern no. & Fluxes\cr

  a & branes: $N_a$ & ($\hat{n}^a_1, \hat{n}^a_2, \hat{n}^a_3$) 
& ( $\hat{m}^a_1, \hat{m}^a_2, \hat{m}^a_3$ ) 
&  $[\frac{(\hat{m}^a_1 + \hat{n}^a_1/2)}{\hat{n}^a_1}, 
 \frac{(\hat{m}^a_2 + \hat{n}^a_2/2)}{\hat{n}^a_2},
\frac{(\hat{m}^a_3 + \hat{n}^a_3/2)}{\hat{n}^a_3}]$ \cr
\hline
Stack-$U_5$ & 5 & $(1, 1, 1)$ & ($-2, -1, 0$) 
& [-$\frac{3}{2}$, -$\frac{1}{2}$,
$\frac{1}{2}$ ]\cr
\hline
Stack-$U_1$ & 1 & $(1, 1, 1)$ & ($-2, 1, 0$) & $[-\frac{3}{2}, \frac{3}{2},
\frac{1}{2} ]$ \cr
\hline
\end{tabular}
\vskip .3cm
\end{center}
\caption{Basic branes for the $SU(5)$ model}
\label{table-1}
\end{table}
Moreover, 
the (chiral) massless spectrum under the resulting gauge group $U(5) \times U(1)$ is summarized in 
Table~\ref{table-2}. The intersection of $U_5$ with $U_1$ is non-chiral since $I_{U_5U_1}$ vanishes.
The corresponding non-chiral massless spectrum shown in the table consists of four pairs of 
${\bf 5}+{\bf{\bar 5}}$ and will be discussed in section~\ref{non-chiral-spectrum}.
\begin{table}[h]
\begin{center}
\begin{tabular}{|l|l|}
\hline
$SU(5)\times  U(1)^2$  & number\\
\hline
$({\bf 10};2,0)$ &   $\ \ 3$ \\
\hline
$({\bf 5};1,1)$ &  $-3$ \\
$(\overline{\bf 5};-1,1)$ &  $4-4$ \\
\hline
\end{tabular}
\end{center}
\caption{Massless spectrum}
\label{table-2}
\end{table}

\subsection{Non-chiral stacks}\label{nonchiral}

So far, we have obtained the gauge and matter chiral spectrum of the $SU(5)$ GUT
using two stacks of magnetized branes. However, in order to complete the model
and stabilize all moduli, one needs to add additional stacks of magnetized branes.
This has to be done in a manner such that the supersymmetries of 
all the brane stacks are mutually compatible. To this end,
we first examine whether the first two stacks $U_5$ and $U_1$ can have 
mutually compatible supersymmetry in a way suitable for moduli
stabilization. The K\"ahler moduli stabilization conditions are written 
in eqs. (\ref{susy}) and (\ref{stab:gen:cond_kahlerX}), corresponding to the cases where 
charged scalar VEVs are respectively zero or non-zero.

Since the VEV of any charged scalar on the $U_5$ stack is 
required to be zero, in order to preserve the gauge symmetry,
the supersymmetry conditions for the $U_5$ stack read:
\begin{equation}
\frac{3}{8} -\frac{1}{2}(J_1 J_2 - 3 J_2J_3 - J_1J_3) = 0,
\label{susy-u5}
\end{equation}
\begin{equation}
	J_1 J_2 J_3 - \frac 14 ( - J_1 - 3 J_2 + 3 J_3) > 0,
\label{+ve-u5} 
\end{equation}
where we have used the fact that all windings are equal to unity 
and that eventually the K\"ahler moduli are stabilized 
according to our ansatz (\ref{oblique-kahler}), such that 
$J_{i\bar{j}} = 0$ for $i\neq j$, and
we have also defined 
\be
	J_{i\bar{i}} \equiv J_i .
\label{diagonal-kahler}
\ee

For the $U_1$ stack on the other hand, one has the option of
turning on a charged scalar VEV without breaking $SU(5)$ 
gauge invariance. However, since all windings are equal to unity,
there are no charged states under $U(1)$ which are $SU(5)$ singlets.
Indeed, there is no antisymmetric representation for $U(1)$, while symmetric 
representations are absent because of our winding choice. The only charged
states then come from the intersection of $U_1$ with $U_5$ (or its image). 
Thus, 
the supersymmetry condition 
for the $U_1$ stack follows from eq. (\ref{susy}), with the 
fluxes given in eq. (\ref{flux-u1}) and Table~\ref{table-1}:
\begin{equation}
- \frac{9}{8} -\frac{1}{2}(J_1 J_2 - 3 J_2J_3 + 3 J_1J_3) = 0,
\label{susy-u1}
\end{equation}
\begin{equation}
	J_1 J_2 J_3 - \frac 14 ( 3 J_1 - 3 J_2 - 9 J_3) > 0.
\label{+ve-u1} 
\end{equation}
Subtracting eq. (\ref{susy-u1}) from eq. (\ref{susy-u5}) one
obtains: $J_1J_3 = - \frac 34$ which is clearly not allowed. We then 
conclude that the $U_1$ stack is not suitable for closed 
string moduli stabilization without charged scalar VEVs from its
intersection with other brane stacks (besides $U_5$). We therefore need
eight new $U(1)$ stacks 
for stabilizing all the nine geometric 
K\"ahler moduli, in the absence of open string VEVs. 

In order to find such new stacks, 
one needs to impose the condition that any chiral fermions, other than 
those discussed in section~\ref{su5gut}, are $SU(5)$ singlets and thus
belong to the `hidden sector', satisfying:
\be
	I_{U_5 a} + I_{U_5 a^*} = 0,\,\,\,\, {\rm for}\,\,a=1,..,8\, .
\label{non-chiral}
\ee
We then introduce eight new stacks $O_1,\dots , O_8$,
which carry in general both oblique and diagonal fluxes in order to stabilize
eight of the geometric K\"ahler moduli, using the supersymmetry
constraints (\ref{susy}). The remaining one is stabilized by the stack $U_5$.
More precisely, to determine the brane stacks $O_1,\dots , O_8$, we start with 
our ansatz for both K\"ahler and complex structure moduli,
and use them to find out the allowed fluxes, consistent with zero net 
chirality and supersymmetry. Later on, we use the resulting fluxes to
show the complete stabilization of moduli, and thus prove the validity 
of our ansatz. 

In general, along a stack-$a$,
the fluxes can be denoted by $3\times 3$ Hermitian 
matrices, 
\begin{eqnarray}
	F^{a}_{(1, 1)} = \begin{pmatrix}f_1 & a & b \cr
	a^* & f_2 & c\cr
	b^*& c^* & f_3 \end{pmatrix},
\label{general-O}
\end{eqnarray}
with $f_i$'s being real numbers, and we have suppressed the 
superscript `$a$'
on the matrix components in the rhs of eq.~(\ref{general-O}). 
The relationships between the 
matrix elements $(F^{a}_{(1, 1)})_{i\bar{j}}$ and the flux components 
$p^{a}_{x^ix^j}$, $p^{a}_{x^iy^j}$, $p^{a}_{y^iy^j}$ are:
\be
	f_i = p_{x^iy^i}\, ,\quad a = p_{x^1y^2} + i p_{x^1x^2}\, ,\quad
	b = p_{x^1y^3} + i p_{x^1x^3}\, ,\quad
	c = p_{x^2y^3} + i p_{x^2x^3}\, . 
\label{fabc}
\ee
The subscript $(1, 1)$ will also sometimes be suppressed for notational simplicity. 
We now solve the non-chirality condition (\ref{non-chiral}) that a general flux of the type (\ref{general-O}) 
must satisfy:
\be
	I_{U_5a} + I_{U_5a^*} = 
	\det (F^{U_5} - F^{a}) + \det (F^{U_5} + F^{a}) = 0\, .  
\label{chiral0}
\ee
The general solution for the flux (\ref{general-O}) is:
\be
	\frac 34 + (f_1 f_2 - 3 f_2 f_3 - f_1 f_3)
			+ (3 c c^*  - a a^* + b b^* ) = 0.
\label{chiral0-1}
\ee
All additional stacks, including $O_1,\dots , O_8$, 
are required to satisfy this condition.

\subsection{Supersymmetry constraint}\label{susy-constraint}

We now impose an additional requirement on the fluxes
along the stacks $O_1,\dots , O_8$, that together with the
stack $U_5$ they should satisfy the supersymmetry conditions
(\ref{susy}), in the absence of charged scalar VEVs. 
Using $F^{a}$ of eq. (\ref{general-O}), the supersymmetry
equations analogous to (\ref{susy-u5}) and (\ref{+ve-u5})
for a stack $O_a$ read:
\begin{eqnarray}
\label{susy-O}
\!\hskip -1cm\left( f_1 f_2 f_3 - c c^* f_1 - bb^* f_2 - aa^* f_3 + a^*bc^* + a b^* c\right) 
-(J_1 J_2 f_3 + J_2J_3 f_1 + J_1J_3 f_2) \!\! &=&\!\! 0,\\ [4pt] 
	J_1 J_2 J_3 - \left[ J_1 (f_2f_3 - cc^*) + J_2 (f_3f_1 - bb^*) 
+ J_3 (f_1f_2 - aa^*)\right]\!\! &>&\!\! 0.
\label{+ve-O} 
\end{eqnarray}

Next, we obtain two sets of fluxes of the form (\ref{general-O})
which satisfy eqs. (\ref{chiral0-1}) and (\ref{susy-O}).
The two sets, $O_1,\dots ,O_4$ and $O_5,\dots ,O_8$,  
are characterized by the diagonal entries in 
the matrix $F^{a}$ (\ref{general-O}), 
which will be the same for the branes of each set. The motivation
behind such choices is dictated by the fact that once 
the off diagonal components of $J_{i\bar{j}}$ are fixed to zero, 
these two sets of fluxes along the diagonal, 
together with the flux of $U_5$ stack, determine the three diagonal
elements $J_i$ (\ref{diagonal-kahler}), completely.

\subsection{Solution for the stacks $O_1,\dots ,O_4$}\label{solutions}

In order to find a constraint on the flux components $f_1, f_2, f_3$ and $a, b, c$
arising out of the requirement that equations (\ref{susy-u5}) and 
(\ref{susy-O}) should be satisfied simultaneously, 
we start with a particular one-parameter
solution of eq. (\ref{susy-u5}):
\begin{equation}
J_1 = \frac{3}{4\epsilon^2},\;\;\;
J_2 = \frac{1}{2\epsilon} + \frac{1}{2},\;\;\;
J_3 = \frac{1}{2\epsilon} - \frac{1}{2}
\label{ji-epsilon}
\end{equation}
for arbitrary parameter $\epsilon\in (0,1)$.\footnote{One can also write down 
a full two-parameter solution of eq.~(\ref{susy-u5}), however we prefer
to use two different one-parameter families with appropriate parametrization
for convenience in model building. The second one-parameter solution will be used in
section~\ref{additional}.} 
Then, by inserting (\ref{ji-epsilon}) into eq. (\ref{susy-O}), one obtains the relation:
\begin{eqnarray}
\frac{3}{4\epsilon^3} (\frac{f_2 + f_3}{2}) 
\!\!\! &+&\!\!\! \frac{1}{4\epsilon^2} [\frac{3}{2}(f_3 - f_2) + f_1]\nonumber\\ 
\!\!\! &=&\!\!\! \left( f_1 f_2 f_3 - c c^* f_1 - bb^* f_2 - aa^* f_3 + a^*bc^* + a b^* c\right)
	+ \frac{f_1}{4}\, .
\label{susy-O1}
\end{eqnarray}

In  solving  eqs. (\ref{chiral0-1}) and (\ref{susy-O1}), 
satisfying also the positivity condition (\ref{+ve-O}), we have to 
keep in mind that $f_i$'s take half-integer values due to the 
NS-NS $B$-field background (\ref{diagonalB}).
On the other hand the parameters $a, b, c$ must be integers, since the 
windings are all one and there is no $B$-field turned on along any 
off-diagonal 2-cycle. 
Our approach is then to 
first look for a solution of eq. (\ref{chiral0-1}) and then 
examine whether such a solution gives an $\epsilon$ 
from eq. (\ref{susy-O1}) such that 
all the $J_i$'s in eq. (\ref{ji-epsilon}) are 
positive. In addition, both positivity 
conditions (\ref{+ve-u5}) and (\ref{+ve-O}) have to be satisfied.

To solve eq.~(\ref{chiral0-1}), we impose the relation
$f_2 = - f_3$. The two equations (\ref{chiral0-1}) and
(\ref{susy-O1}) are then reduced to 
\be
	\frac 34 + 2 f_1f_2 + 3 f_2^2 + 3cc^* + bb^* - aa^* = 0,
\label{chiral0-2}
\ee
and
\begin{equation}
	\frac{1}{4\epsilon^2} ( - 3 f_2 + f_1)
=  - f_1 f_2^2 - c c^* f_1 - bb^* f_2 + aa^* f_2 + a^*bc^* + a b^* c
	+ \frac{f_1}{4}.
\label{susy-O2}
\end{equation}

A solution of eq. (\ref{chiral0-2})
with purely real flux components is found to be:
\be
	f_1 = \frac 52\, ,\quad f_2 = \frac 12\, ,\quad f_3 = -\frac 12\, ,\quad 
	a = 4\, ,\quad b= 3\, ,\quad c=1\, .
\label{soln1-flux}
\ee
Moreover, we notice from eqs. (\ref{chiral0-2}), (\ref{susy-O2}) and the identity:
\be
	a^*bc^* + a b^* c = 2 a_1 (b_1 c_1 + b_2 c_2) 
				+ 2 a_2 (b_2 c_1 - b_1 c_2)\, ,	
\label{abc}
\ee
with $a = a_1 + ia_2$, $b = b_1 + i b_2$,
$c = c_1 + i c_2$, that other solutions can be found 
simply by replacing some of the real components of $a, b, c$
by imaginary ones modulo signs, as long as the values of the products
$aa^*$, $bb^*$, $cc^*$, as well as that of $(a^*bc^* + a b^* c)$
remain unchanged. We make use of such choices for canceling 
off-diagonal $D5$-brane tadpoles which for a general flux matrix 
(\ref{general-O}) read (using eq. (\ref{tad5})):
\begin{eqnarray}
Q^{5, a}_{1\bar{1}} \!\!\! &=&\!\!\! (f_2f_3 - cc^*)\, ,\quad
Q^{5, a}_{2\bar{2}} = (f_3f_1 - bb^*)\, , \quad
Q^{5, a}_{3\bar{3}} = (f_1f_2 - aa^*)\, ,\nonumber\\ [4pt] 
Q^{5, a}_{1\bar{2}} \!\!\! &=&\!\!\! (b^*c - a^*f_3)\, , \quad
Q^{5, a}_{2\bar{3}} = (b^*a - c^*f_1)\, , \quad
Q^{5, a}_{3\bar{1}} = (ac - bf_2)\, .
\label{d5tadpole-1}
\end{eqnarray}
Here we have used the 
complex coordinates $z^i, \bar{z}^i$ and the assumption 
that complex structure is eventually stabilized as in eq.~(\ref{complex}).

The result of our analysis above, giving fluxes for the 
brane stacks $O_1,\dots ,O_4$, (including the solution
(\ref{soln1-flux})) is presented in 
Appendix~A, in eqs. (\ref{fo1}), (\ref{fo2}), (\ref{fo3}), (\ref{fo4}). 
In this Appendix, we also show that the net chiral
fermion contribution from the intersection of each of the four stacks 
$O_1,\dots ,O_4$ with $U_5$ (and its image) is zero,
as shown in eqs. (\ref{chiral-o1}), (\ref{chiral-o2}), (\ref{chiral-o3}),
(\ref{chiral-o4}). Oblique tadpoles $Q^5_{1\bar{2}}$,
$Q^5_{2\bar{3}}$,  $Q^5_{3\bar{1}}$ are given in 
eqs. (\ref{otadpole-o1}), (\ref{otadpole-o2}), (\ref{otadpole-o3}),
(\ref{otadpole-o4}) and their cancellations among these branes
is also apparent. This leaves only diagonal $D5$ tadpoles, given 
in eqs. (\ref{dtadpole-o1}), (\ref{dtadpole-o2}), (\ref{dtadpole-o3}),
(\ref{dtadpole-o4}). The fluxes in real basis are given in eqs.
(\ref{rflux-o1}), (\ref{rflux-o2}), (\ref{rflux-o3}), 
(\ref{rflux-o4}). In Table~\ref{table-3}, we summarize
all Chern numbers and windings for the stacks $O_1,\dots ,O_4$,
as well as those for the stacks $O_5,\dots ,O_8$ appearing in the next
subsection.

From eqs.~(\ref{susy-O}) and (\ref{susy-O2}), the
stacks $O_1,\dots ,O_4$ satisfy the supersymmetry condition:
\begin{equation}
\frac{195}{8} - \frac{1}{2}[-J_1 J_2 + 5 J_2 J_3 + J_1 J_3] = 0,
\label{susy-o1/o4}
\end{equation}
for $\epsilon = \frac{1}{10}$ in eq.~(\ref{ji-epsilon}).
The positivity condition (\ref{+ve-O}) for all of them has the following final form:
\begin{equation}
J_1 J_2 J_3 + \frac{5}{4} J_1 + \frac{41}{4} J_2 +
\frac{59}{4} J_3 > 0,
\label{+ve-o1/o4}
\end{equation}
which is obviously satisfied for the solution (\ref{ji-epsilon})
with $\epsilon = \frac{1}{10}$.
Also, the chiral fermion degeneracies on 
the intersections $U_5-{O_a}$ and  $U_5-{O_a^*}$ are equal to 
\begin{equation}
I_{U_5O_a} = 23\, ,\;\;\;\;I_{U_5O_a^*} = -23\, ,\,\,\,\,\,a=1,\dots ,4\, ,
\label{nonchiral-o1/o4}
\end{equation}
giving vanishing net chirality for all of them individually.
The non-trivial 
tadpole contributions from the four stacks are:
\begin{equation}
Q^9 = 4\, ,\quad Q^5_{x^1y^1} = - 5\, , \quad Q^5_{x^2y^2} = - 41\, , \quad
Q^5_{x^3y^3} = - 59\, .
\label{tadpole-total-o1o4}
\end{equation}

\subsection{Additional stacks: $O_5,\dots ,O_8$}\label{additional}

In the last subsection we found four stacks $O_1,\dots,O_4$
with oblique fluxes but diagonal 5-brane charges. Clearly, in order to stabilize all the
K\"ahler moduli, we need at least four additional stacks with oblique fluxes.
The search for such branes is simplified by observing that
the supersymmetry condition (\ref{susy-u5})  for the stack $U_5$ 
has another one parameter family of solutions, independent of (\ref{ji-epsilon}),
which solves also the condition (\ref{susy-o1/o4}) for the stacks
$O_1,\dots,O_4$:
\begin{eqnarray}
J_1 = \frac{300 \alpha}{4\alpha^2 - 99}\, ,\quad 
J_2 = \alpha\, ,\quad 
J_3 = \frac{99}{4\alpha}\, ,\quad{\rm with}\quad
\alpha^2 > \frac{99}{4}\, .
\label{kahler-2}
\end{eqnarray}

By inserting 
expressions (\ref{kahler-2}) into the general supersymmetry
condition (\ref{susy-O}), and following steps similar to those of
the last subsection, we find the set of stacks 
$O_5,\dots,O_8$ given in Appendix~A, 
with fluxes as in eqs. (\ref{fo5}), (\ref{fo6}), (\ref{fo7}), (\ref{fo8}).
One of these solutions has flux components:
\be
	f_1 = -\frac{25}{2}\, ,\quad f_2 = \frac 12\, ,\quad f_3 = \frac 12\, ,\quad 
	a = -2i\, ,\quad b= -i\, ,\quad c=1\, ,
\label{soln2-flux}
\ee
while the others can be obtained by trivial changes of the off-diagonal elements,
as for the stacks $O_1,\dots,O_4$ discussed in the previous subsection.
Oblique $D5$ tadpoles are written in eqs. 
(\ref{otadpole-o5}), (\ref{otadpole-o6}), (\ref{otadpole-o7}),
(\ref{otadpole-o8}) and the diagonal ones in eqs.
(\ref{dtadpole-o5}), (\ref{dtadpole-o6}), (\ref{dtadpole-o7}),
(\ref{dtadpole-o8}). The net $SU(5)$ non-singlet fermion chirality 
for these stacks is also zero,
as shown in eqs. (\ref{chiral-o5}), (\ref{chiral-o6}), (\ref{chiral-o7}),
(\ref{chiral-o8}). Once again, all off-diagonal $D5$  tadpoles of the type
$Q^5_{1\bar{2}}$, $Q^5_{2\bar{3}}$ and $Q^5_{3\bar{1}}$ cancel
among the contributions of the four brane stacks.
In Table~\ref{table-3}, we summarize the Chern numbers
and windings of the stacks $O_5,\dots ,O_8$, as well.

The four stacks $O_5,\dots ,O_8$ satisfy the supersymmetry condition:
\begin{equation}
\frac{87}{8} - \frac{1}{2}[J_1 J_2 - 25 J_2 J_3 + J_1 J_3] = 0,
\label{susy-o5/o8}
\end{equation}
for
\begin{equation}
\alpha^2 = \frac{99}{4}\times \frac{1431}{1131},
\label{alpha2}
\end{equation}
consistently with the inequality (\ref{kahler-2}). For this value of
$\alpha$, the positivity conditions (\ref{+ve-u5}) and (\ref{+ve-u1})
for the $U_5$ and $U_1$ stacks are also satisfied by $J_i$'s 
of the form (\ref{kahler-2}). On the other hand, using the flux components
(\ref{general-O}) from Table~\ref{table-3},
the positivity condition for the four new stacks takes the following form:
\begin{equation}
J_1 J_2 J_3 + \frac{3}{4} J_1 + \frac{29}{4} J_2 +
\frac{41}{4} J_3 > 0,
\label{+ve-o5/o8}
\end{equation}
and is again obviously satisfied, as is the positivity 
condition (\ref{+ve-o1/o4}) for stacks $O_1,\dots ,O_4$. 
The final uncanceled tadpoles from these stacks are:
\begin{equation}
Q^9 = 4\, ,\quad Q^5_{x^1y^1} = - 3\, ,\quad Q^5_{x^2y^2} = - 29\, ,\quad
Q^5_{x^3y^3} = - 41\, ,
\label{tadpole-total-o5o8}
\end{equation}
while the chiral fermion degeneracy from the intersections
$U_5-O_a$ and $U_5-O_a^*$ is given by:
\begin{equation}
I_{U_5O_a} = 14\, ,\quad I_{U_5O_a^*} = -14\, ,\quad a=5,\dots ,8\, .
\label{nonchiral-o5/o8}
\end{equation}


\begin{table} 
\vskip-0.5cm
\begin{center}
\begin{tabular}{|c||c|c|c|c|c|}
\hline
Stack & No. of & Windings & Diag. Chern no. & 
Diagonal & Oblique\\
&branes:  & $(n^{O_a}_{x^1}, n^{O_a}_{x^2}, n^{O_a}_{x^3})$
&$( m^{O_a}_{x^1y^1}, m^{O_a}_{x^2y^2}, m^{O_a}_{x^3y^3} )$& 
fluxes & Chern no. \\
& $N_{O_a}$ & $(n^{O_a}_{y^1}, n^{O_a}_{y^2}, n^{O_a}_{y^3})$
& & $[f^a_1, f^a_2, f^a_3]$ &  \\

\hline
&&&&&\\
$ O_1$ & $1$& $(1,1,1)$ & (2,0,-1) 
&[$\frac{5}{2}$ ,$\frac {1}{2}$,-$\frac{1}{2}$]& 
$ m^{O_1}_{x^1y^2} = m^{O_1}_{x^2y^1} = 4 $\\
&&$(1,1,1)$&&&$m^{O_1}_{x^1y^3} = m^{O_1}_{x^3y^1} = 3 $\\
&&&&&$ m^{O_1}_{x^2y^3} = m^{O_1}_{x^3y^2} = 1 $\\
\hline
$ O_2$ & $1$& $(1,1,1)$ & (2,0,-1) 
&[$\frac{5}{2}$ ,$\frac {1}{2}$,-$\frac{1}{2}$]&
$ m^{O_2}_{x^1y^2} = m^{O_2}_{x^2y^1} = 4 $\\
&&$(1,1,1)$&&&$m^{O_2}_{x^1y^3} = m^{O_2}_{x^3y^1} = -3$\\
&&&&&$m^{O_2}_{x^2y^3} = m^{O_2}_{x^3y^2} = -1 $\\
\hline
$ O_3$ & $1$& $(1,1,1)$ & (2,0,-1) 
&[$\frac{5}{2}$ ,$\frac {1}{2}$,-$\frac{1}{2}$]& 
$m^{O_3}_{x^1y^2} = m^{O_3}_{x^2y^1} = -4$\\
&&$(1,1,1)$&&&$m^{O_3}_{x^3x^1} = m^{O_3}_{y^3y^1} = 3 $\\
&&&&&$m^{O_3}_{x^2x^3} = m^{O_3}_{y^2y^3} = 1 $\\
\hline
$ O_4$ & $1$& $(1,1,1)$ & (2,0,-1) 
&[$\frac{5}{2}$ ,$\frac {1}{2}$,-$\frac{1}{2}$]& 
$m^{O_4}_{x^1y^2} = m^{O_4}_{x^2y^1} = -4 $\\
&&$(1,1,1)$&&&$m^{O_4}_{x^3x^1} = m^{O_4}_{y^3y^1} = -3 $\\
&&&&&$m^{O_4}_{x^2x^3} = m^{O_4}_{y^2y^3} = -1$\\
\hline
$ O_5$ & $1$& $(1,1,1)$ & (-13,0,0) 
&[$\frac{-25}{2}$ ,$\frac {1}{2}$,$\frac{1}{2}$]& 
$m^{O_5}_{x^1x^2} = m^{O_5}_{y^1y^2} = -2 $\\
&&$(1,1,1)$&&&$m^{O_5}_{x^3x^1} = m^{O_5}_{y^3y^1} = 1 $\\
&&&&&$m^{O_5}_{x^2y^3} = m^{O_5}_{x^3y^2} = 1$\\
\hline

$ O_6$ & $1$& $(1,1,1)$ & (-13,0,0) 
&[$\frac{-25}{2}$ ,$\frac {1}{2}$,$\frac{1}{2}$]& 
$m^{O_6}_{x^1x^2} = m^{O_6}_{y^1y^2} = -2 $\\
&&$(1,1,1)$&&&$m^{O_6}_{x^3x^1} = m^{O_6}_{y^3y^1} = -1 $\\
&&&&&$m^{O_6}_{x^2y^3} = m^{O_6}_{x^3y^2} = -1$\\
\hline
$ O_7$ & $1$& $(1,1,1)$ & (-13,0,0) 
&[$\frac{-25}{2}$ ,$\frac {1}{2}$,$\frac{1}{2}$]& 
$m^{O_7}_{x^1x^2} = m^{O_7}_{y^1y^2} = 2 $\\
&&$(1,1,1)$&&&$m^{O_7}_{x^1y^3} = m^{O_7}_{x^3y^1} = -1 $\\
&&&&&$m^{O_7}_{x^2x^3} = m^{O_7}_{y^2y^3} = 1$\\
\hline

$ O_8$ & $1$& $(1,1,1)$ & (-13,0,0) 
&[$\frac{-25}{2}$ ,$\frac {1}{2}$,$\frac{1}{2}$]& 
$m^{O_8}_{x^1x^2} = m^{O_8}_{y^1y^2} = 2 $\\
&&$(1,1,1)$&&&$m^{O_8}_{x^1y^3} = m^{O_8}_{x^3y^1} = 1 $\\
&&&&&$m^{O_8}_{x^2x^3} = m^{O_8}_{y^2y^3} = -1$\\
\hline

\end{tabular}

\end{center}
\vskip-0.5cm
\caption{Chern numbers and windings of the oblique stacks $O_1,\dots ,O_8$}
\label{table-3}
\end{table}

\subsection{Tadpole cancellation}\label{tadpolecancellation}

We now collect the tadpole contribution from different stacks
to find out how the total RR charges cancel in our model by
adding two extra stacks of single branes, $A$ and $B$.
The tadpole contributions from stacks $O_1,\dots  ,O_4$ 
with oblique fluxes, are given in eq.~(\ref{tadpole-total-o1o4}),
while those from stacks $O_5,\dots  O_8$ are given
in eq.~(\ref{tadpole-total-o5o8}).
In addition, the stacks $U_5$ and $U_1$ together contribute:
\begin{equation}
Q^9 = 6\, ,\quad Q^5_{x^1y^1} = - \frac{1}{2}\, ,\quad
Q^5_{x^2y^2} = - \frac{9}{2}\, ,\quad
Q^5_{x^3y^3} = \frac{3}{2}\, ,
\label{tadpole-total-u5u1}
\end{equation}
where we used the flux components (\ref{flux-u5}) and (\ref{flux-u1}).
These tadpoles are then saturated by the brane stacks 
$A$ and $B$ of Table~\ref{table-4}.
\begin{table} 
\vskip 0.4cm
\begin{center}
\begin{tabular}{|c|c|c|c|c|}
\hline
Stack no. & No. of & Windings  & Chern no. & Fluxes\cr

a & branes: $N_a$& ($\hat{n}^a_1, \hat{n}^a_2, \hat{n}^a_3$) 
& ( $\hat{m}^a_1, \hat{m}^a_2, \hat{m}^a_3$ ) 
&  $[\frac{(\hat{m}^a_1 + \hat{n}^a_1/2)}{\hat{n}^a_1}, 
 \frac{(\hat{m}^a_2 + \hat{n}^a_2/2)}{\hat{n}^a_2},
\frac{(\hat{m}^a_3 + \hat{n}^a_3/2)}{\hat{n}^a_3}]$ \cr
\hline

Stack-A & 1 & $(1, 1, 1)$ & ($147, 0, 0$) & [$\frac{295}{2}$, $\frac{1}{2}$,
$\frac{1}{2}$ ]\cr

\hline

Stack-B & 1 & $(1, 1, 1)$ & ($1, 16, 0$) & $[\frac{3}{2}, \frac{33}{2},
\frac{1}{2} ]$ \cr

\hline

\end{tabular}
\vskip .3cm
\end{center}
\caption{A and B branes}
\label{table-4}
\end{table}
Their contributions to the tadpoles are:
\begin{equation}
Q^9 = 2\, ,\quad Q^5_{x^1y^1} = \frac{34}{4}\, ,\quad
Q^5_{x^2y^2} =  \frac{298}{4}\, ,\quad
Q^5_{x^3y^3} = \frac{394}{4}\, ,
\end{equation}
which precisely cancel the contributions from eqs. 
(\ref{tadpole-total-o1o4}), (\ref{tadpole-total-o5o8})
and (\ref{tadpole-total-u5u1}). Moreover, chiral fermion
multiplicities from the intersections of stacks $A$ and $B$ with $U_5$ vanish, as well:
\begin{equation}
	I_{U_5 A} = I_{U_5 A^*} = I_{U_5 B} = I_{U_5 B^*} =0\, . 
\label{chiral-AB=0}
\end{equation}

We have thus obtained fluxes for the twelve stacks, 
saturating both $D9$ and $D5$ tadpoles. However, for 
supersymmetry, we have only discussed the conditions for nine of the 
twelve brane stacks, namely $U_5$ and $O_1,\dots ,O_8$.
The status of supersymmetry for the brane stacks $U_1$, $A$ and $B$ 
will be studied later, in section~\ref{fiparameter}. 

Before closing this section, we also examine briefly
whether it would be possible to manage tadpole cancellation without adding
the extra stacks $A$ and $B$, within the context
of our construction specified by the choice (\ref{i51,i51'})
of intersection numbers.  Note that the
nine stacks $U_5$ and $O_1,\dots ,O_8$
were the minimal ones needed for K\"ahler moduli stabilization, since the
use of the $U_1$ brane for this purpose was ruled out,
as we discussed in section~\ref{nonchiral}.
The $U_1$ stack on the other hand is needed to get the
right $SU(5)$ particle spectrum. 
Thus, in order to avoid the use of stacks $A$ and $B$, one needs
to examine whether there are solutions,  other than the one found in 
eq. (\ref{flux-u1}), for fluxes along the stack-$U_1$ such that 
tadpole cancellations are possible, while a scalar VEV
charged under this $U(1)$ 
may have to be turned on in order to maintain supersymmetry. 
In such a situation, one needs a winding number three ($\det W = 3$) 
for the stack $U_1$ to saturate the $D9$ tadpole. Moreover, all 
oblique fluxes along the $U_1$ stack have to vanish, otherwise 
they would give rise to uncanceled tadpoles in oblique directions.
Then, by writing the tadpole contributions of three diagonal 
fluxes $f_i$ satisfying the constraint (\ref{i51,i51'}), 
it can be easily seen that one is not able to cancel the combined tadpoles
from stacks $U_5$ and $O_1,\dots ,O_8$. Such a possibility
is therefore ruled out. Of course, one could try to find a solution that
satisfies the constraint (\ref{i51,i51'}) but not necessarily (\ref{i51+i51'}),
as we discussed already in section~\ref{su5gut}. Alternatively,
one can possibly
attempt to manage with just two stacks $U_1$ and $A$,
by using winding number two in one of them. These are straight-forward
exercises for the interested reader who would like to examine these cases.

\subsection{Non-chiral spectrum}\label{non-chiral-spectrum}

The degeneracies of non-chiral states coming from intersections
of the stack $U_5$ with $O_a$ and $O_a^*$ are already given 
in eqs. (\ref{nonchiral-o1/o4}) and (\ref{nonchiral-o5/o8}),
leading to $4\times (23+14)=148$ pairs of $({\bf 5}+{\bf\bar 5})$
representations of $SU(5)$.
They follow from the degeneracy formulae (\ref{scmult'}),
when the net numbers of left- and right-handed fermions are 
equal. In our case, this is insured since
$I_{U_5 O_a} = - I_{U_5 O_a^*}$. 
However, non-chiral particle spectrum
also follows from eqs. (\ref{scmult'}), (\ref{dcmult''}) 
and (\ref{dcmult'}), when any of $I_{ab}$, $I_{ab^*}$, 
$I^A_{aa^*}$ and $I^S_{aa^*}$ are zero, as explained at the end
of section~\ref{spectrum}. This occurs because for instance
$\prod_i (\tilde{\hat{m}}_i^a {\hat{n}}_i^b \pm
{\hat{n}}_i^a  \tilde{\hat{m}}_i^b)$ vanishes
along one or more of the 2-tori, $T^2_j$. 
Similarly for $I^A_{aa^*}$ or $I^S_{aa^*}$, this occurs because of the 
vanishing of fluxes along one or more  of the $T^2$'s.
Given the fluxes in stack $U_5$, which are non-zero
along all three $T^2$'s, non-chiral states can
come only from various intersections of the
$U_5$ stack with other branes. 

For example, the intersection numbers between stacks $U_5$ and
$U_1$ are given in eq.~(\ref{i51,i51'}). One sees that
$I_{U_5 U_1}$ is zero as $(\tilde{\hat{m}}_i^{U_5} 
{\hat{n}}_i^{U_1} -
{\hat{n}}_i^{U_5}  \tilde{\hat{m}}_i^{U_1})$ vanishes
along $T^2_1$ and $T^2_3$. However, in this case there exists
a non-zero intersection number in $d=8$ dimensions corresponding to the $T^2_2$
compactification of the $d=10$ theory, given by:
\begin{equation}
	I_{U_5 U_1}|_{T^2_1, T^2_3} = (\tilde{\hat{m}}_2^{U_5} 
{\hat{n}}_2^{U_1} -
{\hat{n}}_2^{U_5}  \tilde{\hat{m}}_2^{U_1}) = -2,
\label{u5u1-d8}
\end{equation}
with the subscripts $T^2_1, T^2_3$ of $I_{U_5U_1}|$ standing for 
those tori along which the intersection number vanishes. 
This implies two negative chirality (right-handed) fermions in $d=8$, in 
the fundamental representation of $SU(5)$. Under further compactification along 
$T^2_1$ and $T^2_3$, we get four Dirac spinors 
in $d=4$, or equivalently four pairs of $({\bf 5}+{\bf\bar 5})$ Weyl fermions,
shown already in the massless spectrum of Table~\ref{table-2}. They give rise
to four pairs of electroweak higgses, having non-vanishing tree-level Yukawa
couplings with the down-type quarks and leptons, as it can be easily seen.

A similar analysis for the remaining stacks $A$ and $B$ gives
chiral spectra in $d=6$ with degeneracies:
\begin{equation}
	I_{U_5 A}|_{T^2_3} = (\tilde{\hat{m}}_1^{U_5} 
{\hat{n}}_1^{A} -
{\hat{n}}_1^{U_5}  \tilde{\hat{m}}_1^{A})\times 
(\tilde{\hat{m}}_2^{U_5} 
{\hat{n}}_2^{A} -
{\hat{n}}_2^{U_5}  \tilde{\hat{m}}_2^{A})
= 149\, ,
\label{u5A-d6}
\end{equation}
and 
\begin{equation}
	I_{U_5 A^*}|_{T^2_2} = (\tilde{\hat{m}}_1^{U_5} 
{\hat{n}}_1^{A} +
{\hat{n}}_1^{U_5}  \tilde{\hat{m}}_1^{A})\times 
(\tilde{\hat{m}}_2^{U_5} 
{\hat{n}}_2^{A} +
{\hat{n}}_2^{U_5}  \tilde{\hat{m}}_2^{A}) = 146\, .
\label{u5A-d6'}
\end{equation}
They give rise to $149 + 146=295$ pairs of $({\bf 5}+{\bf\bar 5})$.
Similarly, we obtain for the stack $B$:
\begin{equation}
	I_{U_5 B}|_{T^2_3} = (\tilde{\hat{m}}_1^{U_5} 
{\hat{n}}_1^{B} -
{\hat{n}}_1^{U_5}  \tilde{\hat{m}}_1^{B})\times 
(\tilde{\hat{m}}_2^{U_5} 
{\hat{n}}_2^{B} -
{\hat{n}}_2^{U_5}  \tilde{\hat{m}}_2^{B})
= 51\, ,
\label{u5B-d6}
\end{equation}
and 
\begin{equation}
	I_{U_5 B^*}|_{T^2_1} = (\tilde{\hat{m}}_2^{U_5} 
{\hat{n}}_2^{B} +
{\hat{n}}_2^{U_5}  \tilde{\hat{m}}_2^{B})\times 
(\tilde{\hat{m}}_3^{U_5} 
{\hat{n}}_3^{B} +
{\hat{n}}_3^{U_5}  \tilde{\hat{m}}_3^{B}) = 16\, ,
\label{u5B-d6'}
\end{equation}
leading to $51 + 16=67$ pairs of $({\bf 5}+{\bf\bar 5})$. 
All these non chiral states become massive by displacing appropriately
the branes $A$ and $B$ in directions along the tori $T^2_3$, $T^2_2$ and
$T^2_3$, $T^2_1$, respectively.

In addition to the states above, there are several $SU(5)$ singlets coming
from the intersections among the branes $O_1,\dots ,O_8$, $U_1$, $A$ and $B$.
Since they do not play any particular role in physics concerning our analysis, 
we do not discuss them explicitly here. However, such scalars from the non-chiral
intersections among $U_1$, $A$ and $B$ will be used in section~\ref{fiparameter}
for supersymmetrizing these stacks, by cancelling the corresponding non-zero 
FI parameters upon turning on non-trivial VEVs for these fields. The corresponding 
non-chiral spectrum will be therefore discussed below, in section~\ref{fiparameter}.

\section{Moduli stabilization}\label{moduli-stabilization}

Earlier, we have found fluxes along the nine 
brane stacks $U_5$, $O_1,\dots ,O_8$, given in Tables~\ref{table-1},
\ref{table-2}, \ref{table-3}, \ref{table-4} and in Appendix~\ref{Appendix-A}, 
consistent with our ansatz (\ref{complex}) for the complex
structure and (\ref{oblique-kahler}) for the geometric K\"ahler moduli.
We now prove our ansatz by showing that both $\tau$ and $J$ are 
uniquely fixed to the values 
(\ref{complex}), (\ref{oblique-kahler}) and (\ref{kahler-2}), (\ref{alpha2}).
To show this, we make use of the full supersymmetry conditions
for the $U_5$ stack as well as for the stacks $O_1,\dots, O_8$.

For the complex structure moduli stabilization, we make use of the 
$F^a_{(2,0)}$ condition (\ref{stab:gen:M20_condition}) 
implying that purely holomorphic components of fluxes vanish. 
Then, by inserting the flux components $p_{x^ix^j}$, $p_{x^iy^j}$ $p_{y^iy^j}$,
as given in Table~\ref{table-1} and Table~\ref{table-3},
as well as in Appendix~\ref{Appendix-A}, 
along the $U_5$ and $O_1,..,O_8$ stacks,
we obtain a set of conditions on the complex structure matrix
$\tau$, given explicitly in Appendix~\ref{Appendix-B} in
eqs.~(\ref{1A})~-~(\ref{E14}). These equations imply the 
final answer (\ref{complex}).
The details can be found in Appendix~\ref{Appendix-B}.

For  K\"ahler moduli stabilization, we make use of the 
D-flatness condition in stacks $U_5$, $O_1,\dots O_8$
which amounts to using the last two equations in (\ref{susy}).
Explicit stabilization of the geometric K\"ahler moduli to 
the diagonal form, $J_{i\bar{j}} = 0$, ($i\neq j$) 
is given in eqs.~(\ref{K1})~-~(\ref{V12}) of 
Appendix~\ref{Appendix-C}. For the 
stabilization of the diagonal components, the relevant 
equations are: (\ref{susy-u5}), (\ref{+ve-u5}),
(\ref{susy-o1/o4}), (\ref{+ve-o1/o4}), 
(\ref{susy-o5/o8}), (\ref{+ve-o5/o8}). The final solution
for the stabilized moduli is given in eqs.~(\ref{kahler-2})
and (\ref{alpha2}). The numerical values of $J_i$'s can also be 
approximated as:
\be
J_1 \sim 63.96\, ,\quad J_2\sim 5.59\, ,\quad
J_3 \sim 4.42\, .
\label{kahler-values}
\ee

\section{Supersymmetry of stacks $U_1$, $A$ and $B$ }\label{fiparameter}


We now discuss the supersymmetry of the remaining stacks $U_1$,
$A$ and $B$ by making use of the D-flatness conditions 
(\ref{dterm}), (\ref{stab:gen:cond_kahlerX})
and (\ref{stab:gen:cond_pos}). From these equations, 
suppressing the superscript $a$, we obtain the 
FI parameters $\xi$ 
as: 
\be
 {\xi} = \frac{F^3_{(1, 1)} - J^2 F_{(1, 1)}}{J^3 - J F^2_{(1, 1)}}\, ,
\label{xi/G}
\ee
where we have made use of eq. (\ref{matrixF}) and the canonical 
volume normalization (\ref{ft1}).
Then, using the values of the magnetic fluxes in stacks $U_1$,
$A$ and $B$ from Tables~\ref{table-1} and \ref{table-4}, the explicit
form of the FI parameters in terms of the moduli $J_i$ 
(that are already completely fixed to the values (\ref{kahler-values})) is
given by:
\bea
 {\xi^{U_1}} &=& 
	\frac{-\frac{9}{8} -\frac{1}{2}(J_1 J_2 - 3 J_2J_3 + 3 J_1J_3) }
{ J_1 J_2 J_3 - \frac 14 ( 3 J_1 - 3 J_2 - 9 J_3)}\, ,
\label{xi/G-u1}\\ [10pt]
 {\xi^{A}} &=& 
	\frac{\frac{295}{8} -\frac{1}{2}(J_1 J_2 + 295 J_2J_3 +  J_1J_3) }
{ J_1 J_2 J_3 - \frac 14 ( J_1 + 295 J_2 + 295 J_3)}\, ,
\label{xi/G-uA}\\ [10pt]
 {\xi^{B}} &=& 
	\frac{\frac{33}{8} -\frac{1}{2}(J_1 J_2 + 3 J_2J_3 + 33 J_1J_3) }
{ J_1 J_2 J_3 - \frac 14 ( 33 J_1 + 3 J_2 + 99 J_3)}\, ,
\label{xi/G-uB}
\eea
leading to the numerical values:
\begin{equation}
 {\xi^{U_1}} \sim -0.366\, ,\quad    
 {\xi^{A}}	  \sim -4.753\, ,\quad
 {\xi^{B}}	  \sim -5.173\, .
\label{xi-values}
\end{equation}

On the other hand, the charged scalar VEVs $v_\phi$ entering in the modified
D-flatness conditions (\ref{dterm}) and (\ref{stab:gen:cond_kahlerX}) are related
to the modified FI parameters ${\xi^a}/{G^a}$, as it can be easily seen from 
the expressions (\ref{xioverg2}) and (\ref{gaugecoupling}), that are also relevant
for the perturbativity criterion: $v_\phi<<1$ in string units. Their knowledge
needs determination of the matter field metric $G^a$ on the branes
$U_1$, $A$ and $B$. For this purpose, we make use of 
eq.~(\ref{metric}) with the angles $\theta_i$ defined in eq.~(\ref{theta}).
One finds the following values for the metric $G$ in the three stacks:
\begin{equation}
G^{U_1} \sim 2.815\ ,\quad G^A \sim 50.45\ ,\quad G^B \sim 94.551\, , 
\label{G-values}
\end{equation}
that lead to the modified FI parameters:
\begin{equation}
\frac{\xi^{U_1}}{G^{U_1}} \sim   -0.130\ ,\quad
\frac{\xi^A}{G^A} \sim - 0.094\ ,\quad 
\frac{\xi^B}{G^B} \sim - 0.057\, .
\label{xioverG}
\end{equation}
Note that the positivity conditions (\ref{stab:gen:cond_pos}), 
giving positive gauge couplings through eq.~(\ref{gaugecoupling})
for the stacks $U_1$, $A$ and $B$, hold as well. 
These expressions appear also in the FI parameters $\xi^a$ as  
the denominators in the rhs of eqs.~(\ref{xi/G-u1})~-~(\ref{xi/G-uB}).

The last part of the exercise is to cancel the FI parameters  (\ref{xioverG}) with VEVs of specific
charged scalars living on the branes $U_1$, $A$ and $B$, in order to satisfy the D-flatness
condition (\ref{dterm}). For this we first compute the chiral fermion multiplicities on their
intersections:
\begin{equation}
I_{U_1A} = (F^{U_1}-F^A)^3=0\ ,\quad
I_{U_1B} = (F^{U_1}-F^B)^3=0\ ,\quad
I_{AB} = (F^{A}-F^B)^3=0\, .
\label{multiplicity1}
\end{equation}
Since they all vanish, there are equal numbers of chiral and anti-chiral fields in each of these
intersections. In order to determine separately their multiplicities, we follow the method
used in section~\ref{non-chiral-spectrum} and compute:
\begin{equation}
I_{U_1A}|_{T^2_3} = -149\ ,\quad
I_{U_1B}|_{T^2_3} = 45\ ,\quad
I_{AB}|_{T^2_3} = -2336\, .
\label{nc-multiplicity1}
\end{equation}
These correspond to chiral fermion multiplicities in six dimensions generating 
upon compactification to $d=4$ pairs of left- and right-handed
fermions. We also have:
\begin{equation}
I_{U_1A^*} = (F^{U_1}+F^A)^3=292\ ,\quad
I_{U_1B^*} = (F^{U_1}+F^B)^3=0\ ,\quad
I_{AB^*} = (F^{A}+F^B)^3=149 \times 17\, ,
\end{equation}
\label{multiplicity2}
which gives zero net chirality for the $U_1 - B^*$ intersection. Computing
\begin{equation}
I_{U_1B^*}|_{T^2_1} = 18\, ,
\label{nc-multiplicity2}
\end{equation}
one then finds $18$ pairs of left- and right-handed 
fermions in $d=4$ from this intersection.

As a result, we have the following non-chiral fields,
where the superscript refers to the two stacks between which the open 
string is stretched and 
the subscript denotes the charges under the respective $U(1)$'s :  
($\phi^{U_1A}_{+-}$, $\phi^{U_1A}_{-+}$), ($\phi^{U_1B}_{+-}$,
$\phi^{U_1B}_{-+}$),
($\phi^{AB}_{+-}$, $\phi^{AB}_{-+}$), ($\phi^{U_1B^*}_{++}$,
$\phi^{U_1B^*}_{--}$), with fields in the brackets having 
multiplicities $149$, $45$, $2336$ and $18$, respectively.
Restricting only to possible VEVs for these fields,
eq.~(\ref{dterm}) takes the following form for the stacks
$U_1$, $A$ and $B$:
\bea
 \xi^{U_1} /G^{U_1} +  |\phi^{U_1A}_{+-}|^2 - |\phi^{U_1A}_{-+}|^2 + 
|\phi^{U_1B}_{+-}|^2-|\phi^{U_1B}_{-+}|^2 + |\phi^{U_1B^*}_{++}|^2 
- |\phi^{U_1B^*}_{--}|^2 &=& 0\, ,
\label{A}\\ [7pt]
\xi^A /G^A +|\phi^{U_1A}_{-+}|^2 - |\phi^{U_1A}_{+-}|^2 + |\phi^{AB}_{+-}|^2 
- |\phi^{AB}_{-+}|^2 &=& 0\, ,
\label{B}\\ [7pt]
\xi^B /G^B +|\phi^{U_1B}_{-+}|^2 - |\phi^{U_1B}_{+-}|^2 +|\phi^{AB}_{-+}|^2 
- |\phi^{AB}_{+-}|^2 + |\phi^{U_1B^*}_{++}|^2 - |\phi^{U_1B^*}_{--}|^2 &=& 0\, .
\label{C}
\eea
These equations can also be written as:
\bea
 \xi^{U_1} /G^{U_1} + (v^{U_1})^2 = 0 &\Rightarrow& (v^{U_1})^2 = 
-(\xi^{U_1} /G^{U_1})\, ,
\label{D}\\ [5pt]
 \xi^A /G^A + (v^A)^2 = 0 &\Rightarrow& (v^A)^2 = -(\xi^A /G^A)\, ,
\label{E}\\ [5pt]
 \xi^B /G^B + (v^B)^2 = 0 &\Rightarrow& (v^B)^2 = -(\xi^B /G^B)\, ,
\label{F}
\eea
following the notation of eq.~(\ref{stab:gen:cond_kahlerX}), where we defined:
\bea
(v^{U_1})^2 &=& |\phi^{U_1A}_{+-}|^2 - |\phi^{U_1A}_{-+}|^2 + 
|\phi^{U_1B}_{+-}|^2-|\phi^{U_1B}_{-+}|^2 + |\phi^{U_1B^*}_{++}|^2 - 
|\phi^{U_1B^*}_{--}|^2 \nonumber\\ [5pt]
&\equiv& (v^{U_1A})^2+(v^{U_1B})^2+(v^{U_1B^*})^2\, ,
\label{G}\\ [5pt]
(v^A)^2 &=& |\phi^{U_1A}_{-+}|^2 - |\phi^{U_1A}_{+-}|^2 + |\phi^{AB}_{+-}|^2 
- |\phi^{AB}_{-+}|^2 \nonumber\\ [5pt]
&\equiv& -(v^{U_1A})^2+(v^{AB})^2\, ,
\label{H}\\ [5pt]
(v^B)^2 &=& |\phi^{U_1B}_{-+}|^2 - |\phi^{U_1B}_{+-}|^2 +|\phi^{AB}_{-+}|^2 
- |\phi^{AB}_{+-}|^2 + |\phi^{U_1B^*}_{++}|^2 - |\phi^{U_1B^*}_{--}|^2 \nonumber\\ [5pt]
&\equiv& -(v^{U_1B})^2 - (v^{AB})^2+(v^{U_1B^*})^2\, ,
\label{I}
\eea
with for instance $(v^{AB})^2 = |\phi^{AB}_{+-}|^2 - |\phi^{AB}_{-+}|^2$ and 
similarly for the others.

Since we have three equations and four unknowns,
we choose to obtain a special solution by setting $(v^{U_1B})^2 = 0$. 
Equations (\ref{G})~-~(\ref{I}) then give:
\bea
(v^{U_1A})^2+(v^{U_1B^*})^2 &=& -(\xi^{U_1} /G^{U_1}) \sim 0.130\, ,
\label{J}\\
-(v^{U_1A})^2+(v^{AB})^2 &=& -(\xi^A /G^A) \ \, \sim 0.094\, ,
\label{K}\\
- (v^{AB})^2+(v^{U_1B^*})^2 &=&  -(\xi^B /G^B) \ \, \sim 0.057\, ,
\label{L}
\eea
that can be solved to obtain:
\begin{equation}
(v^{U_1A})^2 = -0.011\, , \quad 
(v^{U_1B^*})^2 = 0.141\, , \quad 
 (v^{AB})^2= 0.084\, . 
\label{M}
\end{equation}
Recalling from eqs.~(\ref{G})~-~(\ref{I}) that
\begin{equation}
(v^{U_1A})^2 = |\phi^{U_1A}_{+-}|^2 - |\phi^{U_1A}_{-+}|^2\, , \quad
(v^{U_1B^*})^2 = |\phi^{U_1B^*}_{++}|^2 - |\phi^{U_1B^*}_{--}|^2\, , \quad
(v^{AB})^2 = |\phi^{AB}_{+-}|^2 - |\phi^{AB}_{-+}|^2\, ,
\label{N}
\end{equation}
and comparing with the results of eq.~(\ref{M})
(taking into account the different signs),  
VEVs for the fields $\phi^{U_1A}_{-+}$, $\phi^{U_1B^*}_{++}$and 
$\phi^{AB}_{+-}$  are switched on. Moreover, as required by the validity 
of the approximation, 
the values of the charged scalar VEVs satisfy the condition $v^a << 1$ in string units.



\section{Conclusions}\label{conclusions}

In conclusion, in this work, we have constructed a three generation $SU(5)$
supersymmetric GUT in simple toroidal compactifications of type I string
theory with magnetized $D9$-branes. All 36 closed string moduli are fixed in a
${\cal N}=1$ supersymmetric vacuum, apart from the dilaton, in a way that the
$T^6$-torus metric becomes diagonal with the six internal radii given in
terms of the integrally quantized magnetic fluxes. Supersymmetry requirement
and RR tadpole cancellation conditions impose some of the charged open
string scalars (but $SU(5)$ singlets) to acquire non-vanishing VEVs,
breaking part of the $U(1)$ factors. The rest become massive by absorbing
the RR scalars which are part of the K\"ahler moduli supermultiplets. Thus,
the final gauge group is just $SU(5)$ and the chiral gauge non-singlet
spectrum consists of three families with the quantum numbers of quarks and
leptons, transforming in the ${\bf 10} + {\bf\bar{5}}$ representations of $SU(5)$.
It is of course desirable to study the physics of this model in detail and
perhaps to construct other more `realistic' variations, using the same
framework which has an exact string description. Some of the obvious
questions to examine are:
\begin{enumerate}
\item Give a mass to the non-chiral gauge non-singlet states with the
quantum numbers of higgses transforming in pairs of ${\bf 5} + {\bf\bar{5}}$
representations, keeping massless only one pair needed to break the
electroweak symmetry. A first partial discussion was given in section~\ref{non-chiral-spectrum}.
\item Break the $SU(5)$ GUT symmetry down to the Standard Model, which can
be in principle realized at the string level separating the $U(5)$ stack
into $U(3)\times U(2)$ by parallel brane displacement. However, one would
like to realize at the same time the so-called doublet-triplet splitting for
the Higgs ${\bf 5} + {\bf\bar{5}}$ pair, i.e. giving mass to the unwanted triplets
which can mediate fast proton decay and invalidate gauge coupling
unification, while keeping the doublets massless. One possibility would be
to deform the model by introducing angles, in realizing the $SU(5)$
breaking. In any case, problems (1) and (2) may be related.
\item Compute and study the Yukawa couplings. A general defect of the
present construction, already known in the literature, is the absence of
up-type Yukawa couplings. In this respect, some recent progress using
$D$-brane instantons may be useful for up-quark mass 
generation\cite{March,cvetic,Billo:2007sw}.
\item Study the question of supersymmetry breaking. An attractive direction
would be to start with a supersymmetry breaking vacuum in the absence of
charged scalar VEVs for the extra branes needed to satisfy the RR tadpole
cancellation, $U(1)\times U(1)_A\times U(1)_B$ in our construction. This
`hidden sector' could then mediate supersymmetry breaking, which is mainly
of D-type, to the Standard Model via gauge interactions. Gauginos can then
acquire Dirac masses at one loop without breaking the R-symmetry, due to the
extended supersymmetric nature of the gauge sector~\cite{Antoniadis:2006eb}.
\end{enumerate}
Thus, this framework offers a possible self-consistent setup for string
phenomenology, in which one can build simple calculable models of particle
physics with stabilized moduli and implement low energy supersymmetry
breaking that can be studied directly at the string level.

\section*{Acknowledgments}
This research project has been supported in part by the European Commission under the RTN contract MRTN-CT-2004-503369, in part by a Marie Curie Early Stage
Research Training Fellowship of the European Community's Sixth Framework
Programme under contract number (MEST-CT-2005-0020238 - EUROTHEPHY) and
in part by the INTAS contract 03-51-6346. 

\appendix
\section{
Explicit solutions for $O_1,\dots ,O_8$}\label{Appendix-A}

In this Appendix, we write all the fluxes in the complex coordinate basis 
$(z,{\bar z})$ with $z = x + i y$. 
Then, for the windings and 
1st Chern numbers of Table~\ref{table-1}, we obtain:
\begin{eqnarray}
F^{U_5}_{(1, 1)} = 
-\frac{i}{2} \begin{pmatrix} dz_1 & dz_2 &dz_3 \end{pmatrix}
\begin{pmatrix}-\frac{3}{2} & & \cr
	& -\frac{1}{2} & \cr
	& & \frac{1}{2} \end{pmatrix}
\begin{pmatrix} d\bar{z}_1 \cr d\bar{z}_2 \cr d\bar{z}_3  
\end{pmatrix} . 
\end{eqnarray}
Below, we sometimes suppress the subscript $(1, 1)$ 
to keep the expressions simpler.
The fluxes of the 8 stacks $O_1,\dots ,O_8$ can also be
written in the same coordinate basis:
\begin{eqnarray}
F^{O_1}_{(1, 1)} = 
-\frac{i}{2} \begin{pmatrix} dz_1 & dz_2 &dz_3 \end{pmatrix}
\begin{pmatrix}\frac{5}{2} & 4 & 3\cr
     4	& \frac{1}{2} & 1 \cr
     3	& 1 & -\frac{1}{2} \end{pmatrix}
\begin{pmatrix} d\bar{z}_1 \cr d\bar{z}_2 \cr d\bar{z}_3  
\end{pmatrix}.
\label{fo1}
\end{eqnarray}
From eq. (\ref{fo1}) we get 
\begin{equation}
|F^{U_5} + F^{O_ 1}| = 23\, ,\quad |F^{U_5} - F^{O_1}| = -23\, ,\quad
|F^{O_1}| = \frac{195}{8}\, ,
\label{chiral-o1}
\end{equation}
where we have used the notation 
$|F^{U_5} + F^{O_1}| \equiv \det(F^{U_5} + F^{O_1})$ etc.
The oblique $D5$ tadpoles are:
\begin{equation}
Q^{O_1}_{1\bar{2}} = 3+2\, ,\quad Q^{O_1}_{2\bar{3}} = 12 - \frac{5}{2}\, ,\quad
Q^{O_1}_{3\bar{1}} = 4 - \frac{3}{2}\, ,
\label{otadpole-o1}
\end{equation}
while the diagonal ones are:
\begin{equation}
Q^{O_1}_{1\bar{1}} = -\frac{5}{4}\, ,\quad Q^{O_1}_{2\bar{2}} =  - \frac{41}{4}\, ,\quad
Q^{O_1}_{3\bar{3}} = - \frac{59}{4}\, .
\label{dtadpole-o1}
\end{equation}
In real coordinates, the fluxes are:
\begin{equation}
p^{O_1}_{x^1y^1} = \frac{5}{2},\; p^{O_1}_{x^2y^2} =-p^{O_1}_{x^3y^3} = \frac{1}{2},\;
p^{O_1}_{x^1y^2} = p_{x^2y^1} = 4,\;
p^{O_1}_{x^1y^3} = p^{O_1}_{x^3y^1} = 3,\;
p^{O_1}_{x^2y^3} = p^{O_1}_{x^3y^2} = 1.
\label{rflux-o1}
\end{equation}

The 1st Chern numbers given in Table~\ref{table-4} can then be
read directly from the values of fluxes given above. 
We now give similar data for the stacks $O_2,\dots ,O_8$:
\begin{eqnarray}
F^{O_2}_{(1, 1)} = 
-\frac{i}{2} \begin{pmatrix} dz_1 & dz_2 &dz_3 \end{pmatrix}
\begin{pmatrix}\frac{5}{2} & 4 & - 3\cr
     4	& \frac{1}{2} & - 1 \cr
     - 3  & - 1 & -\frac{1}{2} \end{pmatrix}
\begin{pmatrix} d\bar{z}_1 \cr d\bar{z}_2 \cr d\bar{z}_3  
\end{pmatrix},
\label{fo2}
\end{eqnarray}
leading to: 
\begin{equation}
|F^{U_5} + F^{O_2}| = 23\, ,\quad |F^{U_5} - F^{O_2}| = -23\, ,\quad
|F^{O_2}| = \frac{195}{8}\, .
\label{chiral-o2}
\end{equation}
The oblique tadpoles are:
\begin{equation}
Q^{O_2}_{1\bar{2}} = 3+2\, ,\quad Q^{O_2}_{2\bar{3}} = -12 + \frac{5}{2}\, ,\quad
Q^{O_2}_{3\bar{1}} = - 4 + \frac{3}{2}\, ,
\label{otadpole-o2}
\end{equation}
while the diagonal tadpoles are:
\begin{equation}
Q^{O_2}_{1\bar{1}} = -\frac{5}{4}\, ,\quad Q^{O_2}_{2\bar{2}} =  - \frac{41}{4}\, ,\quad
Q^{O_2}_{3\bar{3}} = - \frac{59}{4}\, .
\label{dtadpole-o2}
\end{equation}
The fluxes in the real basis are:
\begin{equation}
p^{O_2}_{x^1y^1} = \frac{5}{2},\; p^{O_2}_{x^2y^2} =-p^{O_2}_{x^3y^3} = \frac{1}{2},\;
p^{O_2}_{x^1y^2} = p^{O_2}_{x^2y^1} = 4,\;
p^{O_2}_{x^1y^3} = p^{O_2}_{x^3y^1} = -3,\;
p^{O_2}_{x^2y^3} = p^{O_2}_{x^3y^2} = -1.\;
\label{rflux-o2}
\end{equation}
\begin{eqnarray}
F^{O_3}_{(1, 1)} = 
-\frac{i}{2} \begin{pmatrix} dz_1 & dz_2 &dz_3 \end{pmatrix}
\begin{pmatrix}\frac{5}{2} & - 4 & -3i\cr
     -4	& \frac{1}{2} & i \cr
     3i & -i & -\frac{1}{2} \end{pmatrix}
\begin{pmatrix} d\bar{z}_1 \cr d\bar{z}_2 \cr d\bar{z}_3  
\end{pmatrix},
\label{fo3}
\end{eqnarray}
leading to
\begin{equation}
|F^{U_5} + F^{O3}| = 23\, ,\quad |F^{U_5} - F^{O3}| = -23\, ,\quad
|F^{O3}| = \frac{195}{8}\, .
\label{chiral-o3}
\end{equation}
The oblique tadpoles are:
\begin{equation}
Q^{O_3}_{1\bar{2}} = -3-2\, ,\quad Q^{O_3}_{2\bar{3}} = -12i + \frac{5i}{2}\, ,\quad
Q^{O_3}_{3\bar{1}} = - 4i + \frac{3i}{2}\, ,
\label{otadpole-o3}
\end{equation}
and the diagonal ones are:
\begin{equation}
Q^{O_3}_{1\bar{1}} = -\frac{5}{4}\, ,\quad Q^{O_3}_{2\bar{2}} =  - \frac{41}{4}\, ,\quad
Q^{O_3}_{3\bar{3}} = - \frac{59}{4}\, .
\label{dtadpole-o3}
\end{equation}
The fluxes in the real basis are:
\begin{equation}
p^{O_3}_{x^1y^1} = \frac{5}{2},\; p^{O_3}_{x^2y^2} =-p^{O_3}_{x^3y^3} = \frac{1}{2},\;
p^{O_3}_{x^1y^2} = p^{O_3}_{x^2y^1} = -4,\;
p^{O_3}_{x^3x^1} = p^{O_3}_{y^3y^1} = 3,\;
p^{O_3}_{x^2x^3} = p^{O_3}_{y^2y^3} = 1.
\label{rflux-o3}
\end{equation}
\begin{eqnarray}
F^{O_4}_{(1, 1)} = 
-\frac{i}{2} \begin{pmatrix} dz_1 & dz_2 &dz_3 \end{pmatrix}
\begin{pmatrix}\frac{5}{2} & - 4 &  3i\cr
     -4	& \frac{1}{2} & - i \cr
     -3i & i & -\frac{1}{2} \end{pmatrix}
\begin{pmatrix} d\bar{z}_1 \cr d\bar{z}_2 \cr d\bar{z}_3  
\end{pmatrix},
\label{fo4}
\end{eqnarray}
leading to
\begin{equation}
|F^{U_5} + F^{O_4}| = 23\, ,\quad |F^{U_5} - F^{O_4}| = -23\, ,\quad
|F^{O_4}| = \frac{195}{8}\, .
\label{chiral-o4}
\end{equation}
The oblique tadpoles are:
\begin{equation}
Q^{O_4}_{1\bar{2}} = -3-2\, ,\quad Q^{O_4}_{2\bar{3}} = 12i - \frac{5i}{2}\, ,\quad
Q^{O_4}_{3\bar{1}} = 4i - \frac{3i}{2}\, ,
\label{otadpole-o4}
\end{equation}
and the diagonal tadpoles are:
\begin{equation}
Q^{O_4}_{1\bar{1}} = -\frac{5}{4}\, ,\quad Q^{O_4}_{2\bar{2}} =  - \frac{41}{4}\, ,\quad
Q^{O_4}_{3\bar{3}} = - \frac{59}{4}\, .
\label{dtadpole-o4}
\end{equation}
The fluxes in the real basis are:
\begin{equation}
p^{O_4}_{x^1y^1}\! =\! \frac{5}{2},\; p^{O_4}_{x^2y^2} =-p^{O_4}_{x^3y^3}\! =\! \frac{1}{2},\;
p^{O_4}_{x^1y^2} = p^{O_4}_{x^2y^1}\! =\! -4,\;
p^{O_4}_{x^3x^1} = p^{O_4}_{y^3y^1}\! =\! -3,\;
p^{O_4}_{x^2x^3} = p^{O_4}_{y^2y^3}\! =\! -1.
\label{rflux-o4}
\end{equation}








The stacks $O_1,\dots ,O_4$, 
given above, satisfy the supersymmetry conditions 
(\ref{susy-o1/o4}). We now give the set of four stacks,
$O_5,\dots ,O_8$, which satisfy the supersymmetry condition
(\ref{susy-o5/o8}) for the values of $J_i$ given in 
eqs.~(\ref{kahler-2}), (\ref{alpha2}):
\begin{eqnarray}
F^{O_5}_{(1, 1)} = 
-\frac{i}{2} \begin{pmatrix} dz_1 & dz_2 &dz_3 \end{pmatrix}
\begin{pmatrix}-\frac{25}{2} & -2i & -i\cr
     2i	& \frac{1}{2} & 1 \cr
     i & 1 & \frac{1}{2} \end{pmatrix}
\begin{pmatrix} d\bar{z}_1 \cr d\bar{z}_2 \cr d\bar{z}_3  
\end{pmatrix};
\label{fo5}
\end{eqnarray}
\begin{equation}
|F^{U_5} + F^{O_5}| = 14\, ,\quad |F^{U_5} - F^{O_5}| = -14\, ,\quad
|F^{O_5}| = \frac{87}{8}\, ;
\label{chiral-o5}
\end{equation}
\begin{equation}
Q^{O_5}_{1\bar{2}} = i - i\, ,\quad Q^{O_5}_{2\bar{3}} = 2 + \frac{25}{2}\, ,\quad
Q^{O_5}_{3\bar{1}} = -2i + \frac{i}{2}\, ,
\label{otadpole-o5}
\end{equation}
\begin{equation}
Q^{O_5}_{1\bar{1}} = -\frac{3}{4}\, ,\quad Q^{O_5}_{2\bar{2}} =  - \frac{29}{4}\, ,\quad
Q^{O_5}_{3\bar{3}} = - \frac{41}{4}\, ;
\label{dtadpole-o5}
\end{equation}
\begin{equation}
p^{O_5}_{x^1y^1} = -\frac{25}{2},\; p^{O_5}_{x^2y^2} = 
p^{O_5}_{x^3y^3} = \frac{1}{2},\; p^{O_5}_{x^1x^2} = p^{O_5}_{y^1y^2} = -2,\;
p^{O_5}_{x^3x^1} = p^{O_5}_{y^3y^1} = 1,\;
p^{O_5}_{x^2y^3} = p^{O_5}_{x^3y^2} = 1.\;
\label{rflux-o5}
\end{equation}
\begin{eqnarray}
F^{O_6}_{(1, 1)} = 
-\frac{i}{2} \begin{pmatrix} dz_1 & dz_2 &dz_3 \end{pmatrix}
\begin{pmatrix} -\frac{25}{2} & -2i &  i\cr
     2i	& \frac{1}{2} & - 1 \cr
     -i & -1 & \frac{1}{2} \end{pmatrix}
\begin{pmatrix} d\bar{z}_1 \cr d\bar{z}_2 \cr d\bar{z}_3  
\end{pmatrix};
\label{fo6}
\end{eqnarray}
\begin{equation}
|F^{U_5} + F^{O_6}| = 14\, ,\quad |F^{U_5} - F^{O_6}| = -14\, ,\quad
|F^{O_6}| = \frac{87}{8};
\label{chiral-o6}
\end{equation}
\begin{equation}
Q^{O_6}_{1\bar{2}} = i-i\, ,\quad Q^{O_6}_{2\bar{3}} = -2 - \frac{25}{2}\, ,\quad
Q^{O_6}_{3\bar{1}} = 2i - \frac{i}{2}\, ,
\label{otadpole-o6}
\end{equation}
\begin{equation}
Q^{O_6}_{1\bar{1}} = -\frac{3}{4}\, ,\quad Q^{O_6}_{2\bar{2}} =  - \frac{29}{4}\, ,\quad
Q^{O_6}_{3\bar{3}} = - \frac{41}{4}\, ;
\label{dtadpole-o6}
\end{equation}
\begin{equation}
p^{O_6}_{x^1y^1}\! =\! -\frac{25}{2},\; p^{O_6}_{x^2y^2} = 
p^{O_6}_{x^3y^3}\! =\! \frac{1}{2},\; p^{O_6}_{x^1x^2} = p^{O_6}_{y^1y^2}\! =\! -2,\;
p^{O_6}_{x^3x^1} = p^{O_6}_{y^3y^1}\! =\! - 1,\;
p^{O_6}_{x^2y^3} = p^{O_6}_{x^3y^2}\! =\! -1.
\label{rflux-o6}
\end{equation}
\begin{eqnarray}
F^{O_7}_{(1, 1)} = 
-\frac{i}{2} \begin{pmatrix} dz_1 & dz_2 &dz_3 \end{pmatrix}
\begin{pmatrix}-\frac{25}{2} & 2i & - 1\cr
     -2i  & \frac{1}{2} & i \cr
     -1 & -i & \frac{1}{2} \end{pmatrix}
\begin{pmatrix} d\bar{z}_1 \cr d\bar{z}_2 \cr d\bar{z}_3  
\end{pmatrix};
\label{fo7}
\end{eqnarray}
\begin{equation}
|F^{U_5} + F^{O_7}| = 14\, ,\quad |F^{U_5} - F^{O_7}| = -14\, ,\quad
|F^{O_7}| = \frac{87}{8}\, ;
\label{chiral-o7}
\end{equation}
\begin{equation}
Q^{O_7}_{1\bar{2}} = -i+i\, ,\quad Q^{O_7}_{2\bar{3}} = -2i - \frac{25i}{2}\, ,\quad
Q^{O_7}_{3\bar{1}} = -2 + \frac{1}{2}\, ,
\label{otadpole-o7}
\end{equation}
\begin{equation}
Q^{O_7}_{1\bar{1}} = -\frac{3}{4}\, ,\quad Q^{O_7}_{2\bar{2}} =  - \frac{29}{4}\, ,\quad
Q^{O_7}_{3\bar{3}} = - \frac{41}{4}\, ;
\label{dtadpole-o7}
\end{equation}
\begin{equation}
p^{O_7}_{x^1y^1}\! =\! -\frac{25}{2},\; p^{O_7}_{x^2y^2} = 
p^{O_7}_{x^3y^3}\! =\! \frac{1}{2},\; p^{O_7}_{x^1x^2} = p^{O_7}_{y^1y^2}\! =\!  2,\;
p^{O_7}_{x^1y^3} = p^{O_7}_{x^3y^1}\! =\! - 1,\;
p^{O_7}_{x^2x^3} = p^{O_7}_{y^2y^3}\! =\! 1.
\label{rflux-o7}
\end{equation}
\begin{eqnarray}
F^{O_8}_{(1, 1)} = 
-\frac{i}{2} \begin{pmatrix} dz_1 & dz_2 &dz_3 \end{pmatrix}
\begin{pmatrix}-\frac{25}{2} & 2i &  1\cr
     -2i  & \frac{1}{2} & -i \cr
     1 & i & \frac{1}{2} \end{pmatrix}
\begin{pmatrix} d\bar{z}_1 \cr d\bar{z}_2 \cr d\bar{z}_3  
\end{pmatrix};
\label{fo8}
\end{eqnarray}
\begin{equation}
|F^{U_5} + F^{O_8}| = 14\, ,\quad |F^{U_5} - F^{O_8}| = -14\, ,\quad
|F^{O_8}| = \frac{87}{8}\, ;
\label{chiral-o8}
\end{equation}
\begin{equation}
Q^{O_8}_{1\bar{2}} = -i+i\, ,\quad Q^{O_8}_{2\bar{3}} = 2i + \frac{25i}{2}\, ,\quad
Q^{O_8}_{3\bar{1}} = 2 - \frac{1}{2}\, ,
\label{otadpole-o8}
\end{equation}
\begin{equation}
Q^{O_8}_{1\bar{1}} = -\frac{3}{4}\, ,\quad Q^{O_8}_{2\bar{2}} =  - \frac{29}{4}\, ,\quad
Q^{O_8}_{3\bar{3}} = - \frac{41}{4}\, ;
\label{dtadpole-o8}
\end{equation}
\begin{equation}
p^{O_8}_{x^1y^1} = -\frac{5}{2},\; p^{O_8}_{x^2y^2} = 
p^{O_8}_{x^3y^3} = \frac{1}{2},\; p^{O_8}_{x^1x^2} = p^{O8}_{y^1y^2} =  2,\;
p^{O_8}_{x^1y^3} = p^{O_8}_{x^3y^1} =  1,\;
p^{O_8}_{x^2x^3} = p^{O_8}_{y^2y^3} = - 1.
\label{rflux-o8}
\end{equation}

\section{
Complex structure moduli stabilization}\label{Appendix-B}

For each stack of magnetized $D9$-branes, we have three complex 
conditions for the moduli of the complex structure derived from
eq. (\ref{stab:gen:M20_condition}).\\
From stack-$O_1$ :
\bea
4 \tau_{11} + \frac{1}{2} \tau_{21} +  \tau_{31} &=& 
\frac{5}{2}\tau_{12} +4 \tau_{22} + 3 \tau_{32},
\label{1A}\\ [5pt]
3 \tau_{11} + \tau_{21} - \frac{1}{2}\tau_{31} &=& 
\frac{5}{2}\tau_{13} +4 \tau_{23} + 3 \tau_{33}, 
\label{1B}\\ [5pt]
3 \tau_{12} + \tau_{22} - \frac{1}{2}\tau_{32} &=& 4 \tau_{13} + 
\frac{1}{2} \tau_{23} +  \tau_{33}.
\label{1C}
\eea
From stack-$O_2$ :
\bea
4 \tau_{11} + \frac{1}{2} \tau_{21} -  \tau_{31} &=& \frac{5}{2}\tau_{12} 
+4 \tau_{22} - 3 \tau_{32},
\label{2A}\\ [5pt]
-3 \tau_{11} - \tau_{21} - \frac{1}{2}\tau_{31} &=& \frac{5}{2}\tau_{13} 
+4 \tau_{23} - 3 \tau_{33}, 
\label{2B}\\ [5pt]
-3 \tau_{12} - \tau_{22} - \frac{1}{2}\tau_{32} &=& 4 \tau_{13} 
+ \frac{1}{2} \tau_{23} -  \tau_{33}.
\label{2C}
\eea
From stack-$O_3$ :
\bea
-3\tau_{11}\tau_{32} + \tau_{21} \tau_{32} + 3 \tau_{31}\tau_{12} 
- \tau_{31}\tau_{22} + 4 \tau_{11} -  \frac{1}{2} \tau_{21} + 
\frac{5}{2}\tau_{12} - 4 \tau_{22} &=& 0,
\label{3A}\\ [5pt]
-3\tau_{11}\tau_{33} + \tau_{21} \tau_{33} +3 \tau_{13}\tau_{31} 
- \tau_{31}\tau_{23} +  \frac{1}{2} \tau_{31} + \frac{5}{2}\tau_{13} 
- 4 \tau_{23} -3 &=& 0,
\label{3B}\\ [5pt]
-3\tau_{12}\tau_{33}+ \tau_{22} \tau_{33} + 3 \tau_{13}\tau_{32} 
- \tau_{23}\tau_{32} + \frac{1}{2} \tau_{32} - 4 \tau_{13} 
+ \frac{1}{2} \tau_{23} +1 &=& 0.
\label{3C}
\eea
From stack-$O_4$ :
\bea
3\tau_{11}\tau_{32}- \tau_{21} \tau_{32}-3 \tau_{31}\tau_{12} 
+ \tau_{31}\tau_{22} + 4 \tau_{11} -  \frac{1}{2} \tau_{21} 
+ \frac{5}{2}\tau_{12} - 4 \tau_{22} &=& 0,
\label{4A}\\ [5pt]
3\tau_{11}\tau_{33} - \tau_{21} \tau_{33}- 3 \tau_{13}\tau_{31}
+ \tau_{31}\tau_{23}+\frac{1}{2} \tau_{31} + \frac{5}{2}\tau_{13} 
- 4 \tau_{23} +3 &=& 0,
\label{4B}\\ [5pt]
3\tau_{12}\tau_{33} - \tau_{22} \tau_{33} - 3 \tau_{13}\tau_{32}  
+ \tau_{23}\tau_{32}+ \frac{1}{2} \tau_{32} - 4 \tau_{13} 
+ \frac{1}{2} \tau_{23} - 1 &=& 0.
\label{4C}
\eea
From stack-$O_5$ :
\bea
-2\tau_{11}\tau_{22} - \tau_{11}\tau_{32} + 2 \tau_{21}\tau_{12}
+ \tau_{31}\tau_{12} - \frac{1}{2} \tau_{21} - \tau_{31} 
-\frac{25}{2}\tau_{12} -2 &=& 0,
 \label{5A}\\ [5pt]
-2\tau_{11}\tau_{23}- \tau_{11}\tau_{33}+ 2 \tau_{21}\tau_{13}
+ \tau_{31}\tau_{13}-\tau_{21}- \frac{1}{2} \tau_{31}-
\frac{25}{2}\tau_{13} -1 &=& 0,
 \label{5B}\\ [5pt]
-2\tau_{12}\tau_{23}- \tau_{12}\tau_{33}+ 2 \tau_{22}\tau_{13}
+ \tau_{32}\tau_{13}-\tau_{22}- \frac{1}{2} \tau_{32}+\frac{1}{2}\tau_{23} 
+\tau_{33}  &=& 0.
 \label{5C}
\eea
From stack-$O_6$ :
\bea
-2\tau_{11}\tau_{22} + \tau_{11}\tau_{32} + 2 \tau_{21}\tau_{12}
- \tau_{31}\tau_{12} - \frac{1}{2} \tau_{21} + \tau_{31} 
-\frac{25}{2}\tau_{12} -2 &=& 0,
 \label{6A}\\ [5pt]
-2\tau_{11}\tau_{23} + \tau_{11}\tau_{33}+ 2 \tau_{21}\tau_{13}
- \tau_{31}\tau_{13}+\tau_{21}- \frac{1}{2} \tau_{31}
-\frac{25}{2}\tau_{13} + 1 &=& 0,
 \label{6B}\\ [5pt]
-2\tau_{12}\tau_{23} + \tau_{12}\tau_{33}+ 2 \tau_{22}\tau_{13}
- \tau_{32}\tau_{13}+\tau_{22}- \frac{1}{2} \tau_{32} 
+\frac{1}{2}\tau_{23} -\tau_{33}  &=& 0.
 \label{6C}
\eea
From stack-$O_7$ :
\bea
2\tau_{11}\tau_{22} - 2\tau_{21}\tau_{12}+\tau_{21}\tau_{32} 
- \tau_{22}\tau_{31}-\frac{1}{2} \tau_{21}
-\frac{25}{2}\tau_{12}-\tau_{32} +2 &=& 0,
\label{7A}\\ [5pt]
2\tau_{11}\tau_{23} -  2\tau_{21}\tau_{13}+\tau_{21}\tau_{33}
-\tau_{23}\tau_{31}+\tau_{11}-\frac{1}{2} \tau_{31}
-\frac{25}{2}\tau_{13}-\tau_{33} &=& 0,
\label{7B}\\ [5pt]
2\tau_{12}\tau_{23}-2\tau_{22}\tau_{13}+\tau_{22}\tau_{33}
-\tau_{23}\tau_{32} + \tau_{12}-\frac{1}{2} \tau_{32} 
+ \frac{1}{2}\tau_{23} + 1 &=& 0.
\label{7C}
\eea
From stack-$O_8$ :
\bea
2\tau_{11}\tau_{22} - 2\tau_{21}\tau_{12}-\tau_{21}\tau_{32} 
+ \tau_{22}\tau_{31}-\frac{1}{2} \tau_{21}-
\frac{25}{2}\tau_{12}+\tau_{32} +2 &=& 0,
\label{8A}\\ [5pt]
2\tau_{11}\tau_{23} -  2\tau_{21}\tau_{13}-\tau_{21}\tau_{33}
+\tau_{23}\tau_{31}-\tau_{11}-\frac{1}{2} \tau_{31}
-\frac{25}{2}\tau_{13}+\tau_{33} &=& 0,
\label{8B}\\ [5pt]
2\tau_{12}\tau_{23}-2\tau_{22}\tau_{13}-\tau_{22}\tau_{33}
+\tau_{23}\tau_{32} - 
\tau_{12}-\frac{1}{2} \tau_{32} + \frac{1}{2}\tau_{23} - 1&=& 0.
\label{8C}
\eea

Now, from stack-$O_1$ and  stack-$O_2$ one obtains from
eqs.~(\ref{1A}) and (\ref{2A}):
\begin{equation}
\tau_{31} = 3 \tau_{32}\, ,
\label{F1}
\end{equation}
and 
\begin{equation}
4\tau_{11}+ \frac 12 \tau_{21} = \frac 52 \tau_{12} + 4\tau_{22}\, ;
\label{F1'}
\end{equation}
from eqs. (\ref{1B}) and  (\ref{2B}):
\begin{equation}
3\tau_{11}+  \tau_{21} = 3\tau_{33}\, , 
\label{F2}
\end{equation}
and
\begin{equation}
-  \frac 12 \tau_{31} = \frac 52 \tau_{13} + 4 \tau_{23}\, ;
\label{F2'}
\end{equation}
and from eqs. (\ref{1C}) and (\ref{2C}):
\begin{equation}
3\tau_{12}+ \tau_{22} = \tau_{33}\, , 
\label{F3}
\end{equation}
and
\begin{equation}
- \frac 12 \tau_{32} = 4 \tau_{13} + \frac 12 \tau_{23}\, ;
\label{F3'}
\end{equation}

Similarly, from stack-$O_3$ and stack-$O_4$ one has, by
adding eqs. (\ref{3A}) and (\ref{4A}):
\begin{equation}
4 \tau_{11} - \frac 12 \tau_{21} + \frac 52 \tau_{12} 
- 4 \tau_{22} = 0\, ;
\label{F4}
\end{equation}
by adding eqs. (\ref{3B}) and  (\ref{4B}):
\begin{equation}
\frac 12 \tau_{31} + \frac 52 \tau_{13} - 4 \tau_{23} = 0\, ;
\label{F5}
\end{equation}
and  by adding eqs. (\ref{3C}) and  (\ref{4C}):
\begin{equation}
\frac 12 \tau_{32} - 4 \tau_{13} + \frac 12 \tau_{23} = 0.
\label{F6}
\end{equation}



Use of eqs. (\ref{F3'}) and (\ref{F6}) gives: 
\begin{equation}
\tau_{13} = 0\, ,
\label{E1}
\end{equation}
and
\begin{equation}
\tau_{32} + \tau_{23} = 0\, .
\label{E2}
\end{equation}
Moreover, one has from eqs. (\ref{E1}) and (\ref{F5}):
\begin{equation}
\tau_{31} = 8 \tau_{23}\, ; 
\label{E3}
\end{equation}
from eqs. (\ref{E3}) and (\ref{F1}):
\begin{equation}
3 \tau_{32} = 8 \tau_{23}\, ; 
\label{E4}
\end{equation}
from eqs. (\ref{E4}) and (\ref{E2}):
\begin{equation}
\tau_{32} = \tau_{23} = 0\, ;
\label{E5}
\end{equation}
and from eqs. (\ref{E5}) and (\ref{E3}):
\begin{equation}
\tau_{31} = 0\, . 
\label{E6}
\end{equation}

Similarly, use of eqs. (\ref{F1'}) and (\ref{F4}) implies: 
\begin{equation}
\tau_{21} = 5\tau_{12}\, ,
\label{E7}
\end{equation}
and
\begin{equation}
\tau_{11} = \tau_{22}\, ;
\label{E8}
\end{equation}
while use of eq. (\ref{E8}) in eqs. (\ref{F2}) and (\ref{F3}) gives:
\begin{equation}
3\tau_{11} + \tau_{21} - 3 \tau_{33} = 0\, ,
\label{E9}
\end{equation}
and
\begin{equation}
3\tau_{11} + 9 \tau_{12} - 3 \tau_{33} = 0\, .
\label{E10}
\end{equation}

Eqs. (\ref{E9}) and (\ref{E10}) give:
\begin{equation}
\tau_{21} = 9\tau_{12}\, ,
\label{E11}
\end{equation}
which comparing with eq.~(\ref{E7}) implies:
\begin{equation}
\tau_{21} = \tau_{12} = 0\, .
\label{E12}
\end{equation}
Using the result of eq. (\ref{E12}) into eq. (\ref{E9}) then gives (using also eq. (\ref{E8})),
\begin{equation}
\tau_{11} = \tau_{22} = \tau_{33} \equiv \tau\, .
\label{E13}
\end{equation}
The value of $\tau$ is finally determined from any of the bilinear equations, such 
as eq.~(\ref{3B}) or (\ref{3C}):
\begin{equation}
\tau = i\, .
\label{E14}
\end{equation}





\section{
K\"ahler class moduli stabilization}\label{Appendix-C}

For the stabilization of K\"ahler class,
let us denote for definiteness the volume of the 4-cycles
associated to $ J \wedge J$ as
\begin{equation}
 (J \wedge J)_{i\bar{j}}=V_{i\bar{j}}\, ,
\end{equation}
where the indices $i, \bar{j}$ correspond to the $(1,1)$-cycle
perpendicular to the given 4-cycle. 
In the above notation, the supersymmetry conditions on the 
K\"ahler moduli for the various stacks read as follows :

\noindent From stack-$O_1$ using eq. (\ref{fo1}):
\begin{equation}
\frac{195}{8} - \left[\frac{5}{2}V_{1\bar{1}} + 
\frac{1}{2}V_{2\bar{2}} -  \frac{1}{2}V_{3\bar{3}} + 4 V_{1\bar{2}} 
+4 V_{2\bar{1}} +3 V_{1\bar{3}} + 3 V_{3\bar{1}} + V_{2\bar{3}} 
+ V_{3\bar{2}}  \right] = 0,
\label{K1}
\end{equation}
from stack-$O_2$ using eq. (\ref{fo2}):
\begin{equation}
\frac{195}{8} - \left[\frac{5}{2}V_{1\bar{1}} + \frac{1}{2}V_{2\bar{2}} -  
\frac{1}{2}V_{3\bar{3}} + 4 V_{1\bar{2}} +4 V_{2\bar{1}} - 
3 V_{1\bar{3}} - 3 V_{3\bar{1}} - 
V_{2\bar{3}} - V_{3\bar{2}} \right ] = 0,
\label{K2}
\end{equation}
from stack-$O_3$ using eq. (\ref{fo3}):
\begin{equation}
\frac{195}{8} - \left[\frac{5}{2}V_{1\bar{1}} + \frac{1}{2}V_{2\bar{2}} 
-  \frac{1}{2}V_
{3\bar{3}} - 4 V_{1\bar{2}} - 4 V_{2\bar{1}} - 3iV_{1\bar{3}} +3 i V_{3\bar{1}} +
i V_{2\bar{3}} - i V_{3\bar{2}}  \right] = 0,
\label{K3}
\end{equation}
from stack-$O_4$ using eq. (\ref{fo4}):
\begin{equation}
\frac{195}{8} - \left[\frac{5}{2}V_{1\bar{1}} + \frac{1}{2}V_{2\bar{2}} 
-  \frac{1}{2}V_
{3\bar{3}} - 4 V_{1\bar{2}} - 4 V_{2\bar{1}} 
+ 3iV_{1\bar{3}} -3 i V_{3\bar{1}}  - i V_{2\bar{3}} 
+ i V_{3\bar{2}}  \right] = 0,
\label{K4}
\end{equation}
from stack-$O_5$ using eq. (\ref{fo5}):
\begin{equation}
\frac{87}{8} - \left[\frac{-25}{2}V_{1\bar{1}} + \frac{1}{2}V_{2\bar{2}} 
+ \frac{1}{2}V_ {3\bar{3}} - 2i V_{1\bar{2}} + 2i V_{2\bar{1}} - 
i V_{1\bar{3}} + i V_{3\bar{1}} + V_{2\bar{3}} + V_{3\bar{2}}  \right] = 0,
\label{K5}
\end{equation}
from stack-$O_6$ using eq. (\ref{fo6}):
\begin{equation}
\frac{87}{8} - \left [ \frac{-25}{2}V_{1\bar{1}} + 
\frac{1}{2}V_{2\bar{2}} + \frac{1}{2}V_ {3\bar{3}} - 2i V_{1\bar{2}} 
+ 2i V_{2\bar{1}} + i V_{1\bar{3}} - i V_{3\bar{1}} - V_{2\bar{3}} 
- V_{3\bar{2}} \right ] = 0,
\label{K6}
\end{equation}
from stack-$O_7$ using eq. (\ref{fo7}):
\begin{equation}
\frac{87}{8} - \left[\frac{-25}{2}V_{1\bar{1}} + \frac{1}{2}V_{2\bar{2}} 
+ \frac{1}{2}V_ {3\bar{3}} + 2i V_{1\bar{2}} - 2i V_{2\bar{1}} 
- V_{1\bar{3}} -  V_{3\bar{1}} +i V_{2\bar{3}} -i V_{3\bar{2}} \right ] = 0,
\label{K7}
\end{equation}
from stack-$O_8$ using eq. (\ref{fo8}):
\begin{equation}
\frac{87}{8} - \left[\frac{-25}{2}V_{1\bar{1}} 
+ \frac{1}{2}V_{2\bar{2}} + \frac{1}{2}V_ {3\bar{3}} + 2i V_{1\bar{2}} 
- 2i V_{2\bar{1}} + V_{1\bar{3}} +  V_{3\bar{1}} -i V_{2\bar{3}} 
+i V_{3\bar{2}}  \right] = 0.
\label{K8}
\end{equation}

Now, from  stacks-$O_1$ and $O_2$,
eqs. (\ref{K1}) and (\ref{K2}) give:
\begin{equation}
3 \left ( V_{1\bar{3}} +  V_{3\bar{1}} \right ) +\left ( V_{2\bar{3}} 
+ V_{3\bar{2}}  \right ) = 0;
\label{L1}
\end{equation}
from  stacks-$O_3$ and $O_4$,
eqs. (\ref{K3}) and (\ref{K4})  give:
\begin{equation}
-3i \left ( V_{1\bar{3}} - V_{3\bar{1}} \right ) 
+ i \left ( V_{2\bar{3}}-V_{3\bar{2}} \right ) = 0;
\label{L2}
\end{equation}
from  stacks-$O_5$ and $O_6$,
eqs. (\ref{K5}) and (\ref{K6})  give:
\begin{equation}
-i \left ( V_{1\bar{3}} - V_{3\bar{1}} \right ) + \left ( V_{2\bar{3}} 
+ V_{3\bar{2}}  \right ) = 0;
\label{L3}
\end{equation}
and from  stacks-$O_7$ and $O_8$,
eqs. (\ref{K7}) and (\ref{K8}) give:
\begin{equation}
-\left ( V_{1\bar{3}} +  V_{3\bar{1}} \right ) + i \left ( V_{2\bar{3}}
-V_{3\bar{2}} \right ) = 0.
\label{L4}
\end{equation}

Eq. (\ref{L4}) implies 
\begin{equation}
i \left ( V_{2\bar{3}}-V_{3\bar{2}} \right ) = \left ( V_{1\bar{3}} 
+  V_{3\bar{1}} \right )\, ,
\end{equation}
which leads from eq. (\ref{L1})
\begin{equation}
3i \left ( V_{2\bar{3}}-V_{3\bar{2}} \right ) + \left ( V_{2\bar{3}} 
+ V_{3\bar{2}}  \right ) = 0\, .
\label{L5}
\end{equation}
Similarly, eq.(\ref{L3}) implies 
\begin{equation}
i\left ( V_{1\bar{3}} - V_{3\bar{1}} \right ) = 
\left ( V_{2\bar{3}} + V_{3\bar{2}}  \right )\, ,
\end{equation}
which leads from eq. (\ref{L2})
\begin{equation}
-3\left ( V_{2\bar{3}} + V_{3\bar{2}}  \right ) 
+ i \left ( V_{2\bar{3}}-V_{3\bar{2}} \right ) = 0\, . 
\label{L6}
\end{equation}
Now eqs. (\ref{L5}) and (\ref{L6}) can be solved to give
\begin{equation}
V_{2\bar{3}} + V_{3\bar{2}} = 0,
\label{L7}
\end{equation}
and
\begin{equation}
V_{2\bar{3}} - V_{3\bar{2}} = 0,
\label{L8}
\end{equation}
implying 
\begin{equation}
V_{2\bar{3}} = V_{3\bar{2}} = 0. 
\label{V23}
\end{equation}

Then one has from eq. (\ref{L1})
\begin{equation}
 V_{1\bar{3}} +  V_{3\bar{1}} = 0, 
\label{L9}
\end{equation}
and from eq. (\ref{L2})
\begin{equation}
 V_{1\bar{3}} -  V_{3\bar{1}} = 0, 
\label{L10}
\end{equation}
implying 
\begin{equation}
V_{1\bar{3}} =  V_{3\bar{1}} = 0. 
\label{V13}
\end{equation}
Using the obtained values, eqs. (\ref{K1})~-~(\ref{K3})  give
\begin{equation}
 V_{1\bar{2}} +  V_{2\bar{1}} = 0,
\label{L11}
\end{equation}
while  eqs. (\ref{K7})~-~eq. (\ref{K5})  give
\begin{equation}
 V_{1\bar{2}} - V_{2\bar{1}} = 0,
\label{L12}
\end{equation}
implying 
\begin{equation}
V_{1\bar{2}} = V_{2\bar{1}} = 0.
\label{V12}
\end{equation}

\bibliographystyle{fullsort}
\bibliography{bibliography} 

\end{document}